\documentclass[letterpaper, 10 pt, conference]{ieeeconf}
\IEEEoverridecommandlockouts
\usepackage[table,xcdraw,dvipsnames]{xcolor}
\usepackage{booktabs}
\usepackage{graphicx}
\usepackage{epsfig}
\usepackage{amsmath}

\usepackage{amssymb}
\usepackage{amsthm}
\usepackage{wasysym}
\usepackage{mathrsfs}
\usepackage{bbm}
\usepackage{color}
\usepackage[ruled,vlined,linesnumbered]{algorithm2e}
\usepackage[colorlinks=true,linkcolor=black,anchorcolor=black,citecolor=black,filecolor=black,menucolor=black,runcolor=black,urlcolor=black]{hyperref}
\usepackage{etoolbox}
\usepackage{multirow}
\usepackage{mathtools}

\usepackage{siunitx}
\usepackage{outlines}
\let\labelindent\relax
\usepackage{enumitem}
\setenumerate[1]{label=\arabic*.}
\setenumerate[2]{label=\alph*.}

\usepackage{algorithm2e}
\RestyleAlgo{ruled}

\usepackage{cleveref}
\crefname{equation}{}{}
\crefname{figure}{Fig.}{Figs.}
\crefname{table}{Tab.}{Tabs.}
\crefname{section}{Sec.}{Secs.}
\crefname{algorithm}{Alg.}{Algs.}
\crefname{algocf}{Alg.}{Algs.}
\Crefname{algocf}{Alg.}{Algs.}
\crefname{appendix}{Appx.}{Appxs.}
\crefname{thm}{Thm.}{Thms.}
\crefname{cor}{Cor.}{Cors.}
\crefname{lemma}{Lem.}{Lems.}
\crefname{problem}{Prob.}{Probs.}
\crefname{assumption}{Assum.}{Assums.}

\newtheorem{definition}{Definition}
\newtheorem{remark}[definition]{Remark}
\newtheorem{lemma}[definition]{Lemma}
\newtheorem{proposition}[definition]{Proposition}

\newtheorem{corollary}[definition]{Corollary}

\newtheorem*{problem*}{Problem}

\providecommand{\R}{\ensuremath \mathbb{R}}

\newcommand{\regtext}[1]{\mathrm{\textnormal{#1}}}
\newcommand{\ol}[1]{\overline{#1}}
\newcommand{\ul}[1]{\underline{#1}}
\newcommand{\defemph}[1]{\emph{#1}}

\newcommand{\vc}[1]{\mathbf{#1}}
\newcommand{\lbl}[1]{^{\regtext{#1}}}

\newcommand{\st}{\,\regtext{s.t.}\,}

\newcommand{\zeros}{\mathbf{0}}
\newcommand{\ones}{\mathbf{1}}
\newcommand{\eye}{\mathbf{I}}
\newcommand{\basisvec}{\vc{e}}

\newcommand{\norm}[1]{\left\Vert#1\right\Vert}
\newcommand{\abs}[1]{\left\vert#1\right\vert}

\newcommand{\trans}{^\top}
\newcommand{\inv}{^{-1}}

\newcommand{\sign}{\regtext{sign}}

\DeclareMathOperator*{\argmin}{arg\,min}

\newcommand{\given}{\,\mid\,}
\DeclareMathOperator{\vol}{vol}

\newcommand{\dist}[1]{\mathcal{#1}}
\newcommand{\set}[1]{#1}
\newcommand{\setfun}[1]{\mathcal{#1}}
\newcommand{\zonotope}{\set{Z}}
\newcommand{\zonotopefn}{\setfun{Z}}
\newcommand{\normaldist}{\dist{N}}
\newcommand{\pdf}{f}
\newcommand{\minksum}{\oplus}
\newcommand{\conv}{*}
\newcommand{\zonotenttuple}[4]{#1#3,#4#2}
\NewDocumentCommand{\zonotent}{s >{\SplitArgument{1}{,}}r()}{%
    \IfBooleanTF{#1}{\zonotenttuple{\left(}{\right)}#2}{\zonotenttuple{(}{)}#2}}
\NewDocumentCommand{\relaxedzonotent}{s >{\SplitArgument{1}{,}}r()}{%
    \IfBooleanTF{#1}{\zonotenttuple{\left(}{\right)}#2}{\zonotenttuple{(}{)}#2}\lbl{r}}

\newcommand{\zcenter}{c}
\newcommand{\zgen}{G}
\newcommand{\zgenvec}{g}
\newcommand{\zcoef}{\beta}

\newcommand{\densitypeak}{\pdf\lbl{peak}}
\newcommand{\densitymode}{x^\star}
\newcommand{\wall}{w}
\newcommand{\gaussmean}{\mu}
\newcommand{\gausscov}{\Sigma}
\newcommand{\gaussscale}{\alpha}
\newcommand{\gaussmass}[2][]{\nu_{#1}(#2)}

\newcommand{\liftheight}{\lambda}
\newcommand{\halfwidth}{\ell}

\newcommand{\intmass}{\kappa}

\newcommand{\affmat}{A}
\newcommand{\affvec}{b}

\newcommand{\zmass}{\pi}
\newcommand{\convpdf}{g}

\newcommand{\hsnormal}{a}
\newcommand{\hsoffset}{b}
\newcommand{\nfacets}{J}

\newcommand{\momentmean}{\hat{\gaussmean}}
\newcommand{\momentcov}{\hat{\gausscov}}

\newcommand{\stdnormcdf}{\Phi}
\newcommand{\confradius}{r}
\newcommand{\covroot}{L}
\newcommand{\covrot}{Q}

\newcommand{\tileshare}{v}

\newcommand{\boundcov}{\bar{\gausscov}}
\newcommand{\wallshare}{\eta}
\newcommand{\obstacle}{O}

\newcommand{\discardmass}{\zeta}
\newcommand{\nred}{n\lbl{red}}
\newcommand{\tmplidx}{\mathcal{L}}
\newcommand{\tmplgen}{\zgen\lbl{lrg}}
\newcommand{\tmplcenter}{\zcenter'}
\newcommand{\shapescale}{\chi}
\newcommand{\sizescale}{s}
\newcommand{\redzono}{\zonotope'}
\newcommand{\redpdf}{\pdf'}
\newcommand{\containmat}{M}
\newcommand{\containvec}{d}
\newcommand{\truepdf}{p}
\newcommand{\overpdf}{\bar{\truepdf}}
\newcommand{\insidepdf}{q}
\newcommand{\exitmass}{\delta}
\newcommand{\keepset}{\set{K}}
\newcommand{\noisegain}{B}
\newcommand{\mappedpdf}{\tilde{\truepdf}}
\newcommand{\passpdf}{\check{\truepdf}}
\newcommand{\reachset}{\set{Y}}

\newcommand{\procnoise}{d}

\newcommand{\ctsmat}{A}
\newcommand{\ctsnoisegain}{C}
\newcommand{\ctsnoise}{\dot{\rv{W}}}
\newcommand{\pzinput}{u}

\newcommand{\samplestep}{\Delta t}

\newcommand{\unsafeset}{\set{H}}
\newcommand{\pzcov}{\gausscov\lbl{PZ}}
\newcommand{\hullpdf}{\pdf\lbl{hull}}

\newcommand{\sloshangle}{\vartheta}
\newcommand{\sloshfreq}{\omega}
\newcommand{\sloshdamp}{\zeta_{\regtext{s}}}

\newcommand{\besselroot}{j'_{1}}
\newcommand{\gravity}{g\lbl{grav}}
\newcommand{\cupradius}{R}
\newcommand{\filldepth}{h}

\newcommand{\cupaccel}{a}
\newcommand{\cupcmd}{u}
\newcommand{\acceldist}{d}
\newcommand{\surfheight}{z\lbl{s}}
\newcommand{\surfgain}{\gamma}
\newcommand{\spillmargin}{z\lbl{rim}}
\newcommand{\sloshstepmat}{A}
\newcommand{\sloshstepinput}{B_{\regtext{d}}}
\newcommand{\mismatchset}{\set{E}}

\newcommand{\spillrisk}{p\lbl{max}}
\newcommand{\movetime}{\tau}
\newcommand{\nsteps}{N}
\newcommand{\filldepthlo}{\ul{\filldepth}}
\newcommand{\filldepthhi}{\ol{\filldepth}}
\newcommand{\filldepthref}{\filldepth_0}
\newcommand{\sloshstepmatref}{A_0}
\newcommand{\sloshstepinputref}{B_{\regtext{d},0}}
\newcommand{\mismatchrad}{\Lambda}
\newcommand{\maxabs}{\varrho}
\newcommand{\mismatchbound}{\bar{e}}
\newcommand{\accelstd}{\sigma}
\newcommand{\truncmass}{\nu_3}

\NewDocumentCommand{\zonorm}{O{\zonotope} m}{\norm{#2}_{#1}}

\newcommand{\outsidemass}{\epsilon}
\newcommand{\truncmeanerr}{e_{\gaussmean}}
\newcommand{\trunccoverr}{e_{\gausscov}}

\newcommand{\resultfigure}[2]{\IfFileExists{#2}{\includegraphics[width=#1]{#2}}{\fbox{\parbox{\dimexpr#1-2\fboxsep-2\fboxrule\relax}{\centering\footnotesize upload \detokenize{#2}}}}}
\newcommand{\tablenote}[1]{\par\vspace{2pt}\parbox{\linewidth}{\footnotesize #1}}

\newcommand{\rv}[1]{\mathsf{#1}}

\DeclareMathOperator{\diag}{diag}
\DeclareMathOperator{\rank}{rank}
\newcommand{\expect}{\mathbb{E}}
\newcommand{\invtrans}{^{-\top}}
\newcommand{\dx}{\,dx}

\crefname{definition}{Def.}{Defs.}
\crefname{proposition}{Prop.}{Props.}
\crefname{remark}{Rem.}{Rems.}
\crefname{theorem}{Thm.}{Thms.}
\crefname{corollary}{Cor.}{Cors.}
\crefname{example}{Ex.}{Exs.}
\AtBeginEnvironment{definition}{\crefalias{definition}{definition}}
\AtBeginEnvironment{remark}{\crefalias{definition}{remark}}
\AtBeginEnvironment{lemma}{\crefalias{definition}{lemma}}
\AtBeginEnvironment{proposition}{\crefalias{definition}{proposition}}
\AtBeginEnvironment{assumption}{\crefalias{definition}{assumption}}
\AtBeginEnvironment{example}{\crefalias{definition}{example}}
\AtBeginEnvironment{theorem}{\crefalias{definition}{theorem}}
\AtBeginEnvironment{corollary}{\crefalias{definition}{corollary}}
\AtBeginEnvironment{problem}{\crefalias{definition}{problem}}

\usepackage[
    style=ieee,
    doi=true,
    isbn=false,
    url=false,
    eprint=false,
    backend=biber,
    natbib=true,
    citestyle=numeric-comp
    ]{biblatex}

\bibliography{references}

\title{Zonotents: Densities on Robust Sets Enable Probabilistic Order Reduction}

\author{Shreyas Kousik
\thanks{S. Kousik is with the Woodruff School of Mechanical Engineering, Georgia Tech, Atlanta, GA, USA (e-mail: \texttt{shreyas.kousik@me.gatech.edu}).}
\thanks{Project page: \url{https://zonotents.github.io/}.}
}

\begin{document}

\maketitle

\begin{abstract}
Set propagation reachability analysis typically uses robust sets, which bound a system's reachable states but assign no probability to them.
We propose the \textit{zonotent}, which treats a zonotope (a common robust reachable set representation) as the support of a probability density.
We instantiate zonotents with Gaussian densities truncated to the zonotope, optionally atop a uniform density, and provide the common reachability operations of affine maps and Minkowski sums for these \textit{Gaussian zonotents}.
Unlike a Gaussian, a Gaussian zonotent keeps the hard bounds of a robust set.
Our main contribution, probabilistic order reduction, limits the growth of a zonotope's order (analogous to the size of its computation representation, which typically grows during reachability analysis) during propagation by using the density to discard unlikely parts of the support.
Outer reduction encloses the support, whereas inner and free reductions keep all but a chosen share of the probability.
Because our Minkowski sum is approximate, we bound the probability that the true state ever leaves a sequence of sets by propagating a Gaussian zonotent whose density lies above that of the true state.
Numerical experiments on uncertain linear systems show that Gaussian zonotents match the accuracy of untruncated Gaussians and are less conservative than zonotope reachable sets, including when bounding the probability of spilling during an aggressive slide of a cup of water with unknown fill depth.
\end{abstract}

\section{Introduction}\label{sec: intro}

\suppressfloats[t]
\begin{figure}[t]
    \centering
    \includegraphics[width=\columnwidth]{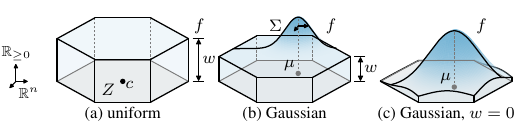}
    \caption{
    A zonotent is a probability density $\pdf$ atop a zonotope, a convex polytope given by a center $\zcenter$ and generator vectors, that is its support $\zonotope$.
    (a) A uniform zonotent, whose density is a constant \textit{wall} of height $\wall$.
    (b) A Gaussian zonotent, a scaled Gaussian with mean $\gaussmean$ and covariance $\gausscov$ atop the wall.
    (c) A Gaussian zonotent with no wall ($\wall = 0$).
    }
    \label{fig: zonotent overview}
    \vspace*{-0.2cm}
\end{figure}

Reachability analysis of uncertain dynamical systems typically propagates compact sets to robustly enclose every reachable state \cite{althoff2021set}.
A commonly-used representation is the zonotope (a convex, centrally symmetric polytope given by an affine map of a hypercube; the affine map is parameterized by a \textit{center} vector and a \textit{generator} matrix), because zonotopes are closed under affine maps and Minkowski sums, the two main operations of set propagation \cite{girard2005reachability}.

However, such robust reachable sets assign no probability to the states they contain, so a planner that certifies safety with them may find no safe motion even when a collision requires an unlikely sequence of worst-case disturbances.
Probabilistic methods instead often propagate Gaussians, sometimes as mixtures \cite{blackmore2011chance,alspach1972nonlinear}, which linear dynamics with Gaussian noise preserve, but since Gaussians are unbounded, they assign probability to states the system can never reach.

Existing combinations of sets and probabilities mostly keep the set and the density apart.
Probabilistic zonotopes \cite{althoff2009safety} bound the means of a set of Gaussians with a zonotope, but not the states.
Truncated Gaussians confine a density to a polytope, but they have been applied mainly to neural network verification \cite{tran2023probstar} and to state constraints in estimation \cite{simon2010kalman}, not to dynamical systems reachability analysis.
We therefore ask whether one representation can keep the compact support of a robust reachable set, assign probability within it, and propagate both by set operations.

Our first key insight is to use the zonotope as the support of a probability density.
We call the pair a \defemph{zonotent}, after the ``tent'' that the density forms atop the zonotope (\Cref{fig: zonotent overview}).
The density of a Gaussian zonotent is a uniform ``wall'' over the zonotope plus a Gaussian truncated to it.
Its support then propagates exactly as a zonotope, and its density propagates through its mean and covariance, exactly under affine maps and approximately under Minkowski sums.
This underlies how zonotents bridge the gap between these representations.

The generator matrix of a zonotope typically grows at every reachability step, and with it the cost of each operation; so, set propagation usually requires \textit{order reduction}, replacing a zonotope by one of lower \textit{order} (generators per dimension) \cite{girard2005reachability,althoff2007reachability}.
Order reduction generally either encloses a zonotope, so the reduced set grows with each reduction, or produces a largest-possible inner subset, but typically with no measure of the states it leaves out.
The probability density gives us just such a measure, giving a new way to simplify a zonotope, which we call \textit{probabilistic order reduction}: a zonotent can discard a part of its support that carries little probability while also reducing its generator count.

We propose three forms of probabilistic order reduction: outer reduction keeps the robust reachable set; inner reduction lies in the support and discards at most a chosen share of the probability, called its budget; and free reduction also discards at most its budget but need not enclose or lie in the original support.
To the best of our knowledge, probabilistic order reduction is the first order reduction of a zonotope that uses a density to choose the reduced support and bounds the discarded probability mass.

\subsubsection*{Contributions}
In summary, we contribute the following:
\begin{enumerate}
    \item We define zonotents and their operations, and instantiate them with Gaussian densities (which includes uniform zonotents) (\Cref{sec: method zonotents,sec: gaussian zonotents,sec: minkowski sum}).
    \item We propose probabilistic order reduction in outer, inner, and free forms (\Cref{sec: order reduction})
    
    \item We bound the change in any probability from probabilistic order reduction, and bound the probability that a system's true state leaves a chain of reduced supports or safe sets (\Cref{sec: first exit,prop: reduction guarantee,prop: free reduction,prop: first exit}).
    \item We validate our zonotent representation and operations in simulation, in comparison to untruncated Gaussians, zonotopes, and probabilistic zonotopes \cite{althoff2009safety} in reachability analysis, most notably the difficult problem of predicting if a sliding cup of water will spill using a linear slosh model, validated against a fluid simulator (\Cref{sec: experiments}).
\end{enumerate}

\noindent Besides those noted above, \Cref{sec: related work} reviews related work, \Cref{sec: preliminaries} gives notation, and \Cref{sec: conclusion} concludes the paper.

\section{Related Work}\label{sec: related work}

We now review related work in robust set propagation, mixed set and Gaussian representations, probabilistic reachable sets, and risk over reachable sets.
To the best of our knowledge, none of these methods keeps a compact reachable support and a density on it in one object, or uses the density to reduce the order of the support.

\subsubsection*{Robust Set Propagation}
Set-propagation reachability analysis computes the set of all states that a system can reach under bounded uncertainty \cite{althoff2021set}.
Zonotopes are closed under affine maps and Minkowski sums \cite{girard2005reachability}; constrained zonotopes extend this closure to intersections \cite{scott2016constrained} and ellipsotopes combine zonotopes with ellipsoids \cite{kousik2023ellipsotopes}.
Each Minkowski sum adds generators to these representations, so outer order reduction encloses these sets in a set with fewer generators \cite{girard2005reachability,kopetzki2017methods}, while inner order reduction finds one inside it \cite{lutzow2025underapproximative}.
Probabilistic order reduction extends these reductions by using the density to choose the reduced support.

\subsubsection*{Mixed Set and Gaussian Representations}

Probabilistic zonotopes \cite[Def.~4]{althoff2009safety} split the generators into bounded and Gaussian classes to represent a set of Gaussians whose means range over a zonotope.
They bound the probability of an unsafe set by integrating an enclosing hull of these Gaussians over it.
Zonotopic Kalman filters choose Kalman-like gains that minimize the size of a zonotope \cite{combastel2015zonotopes}, and an extension propagates a zonotope for bounded errors together with a Gaussian for stochastic ones \cite{combastel2016extended}.
In both, the set and the Gaussian add, so the bounded part is enclosed exactly but carries no distribution, and the state has unbounded support.
The zonotopic mixture filter \cite{gonzalez2026zonotopic} runs such a filter for each history of the modes of bounded noise and weights the zonotope of each history by its probability, so its probabilities lie on the modes rather than on the states within each zonotope.
It limits the number of zonotopes by merging some into enclosing ones, which, like outer order reduction, keeps every state consistent with the measurements.
Probabilistic affine forms attach bounds on cumulative distributions or concentration inequalities to the noise symbols of a zonotope \cite{bouissou2012generalization,bouissou2016uncertainty}, which bound probabilities of sets without a density on the support.
ProbStar instead truncates a Gaussian on the predicate variables of a star set, which is an affine image of a polytope of these variables called the predicate, to that polytope and propagates it exactly through the affine and piecewise-affine layers of neural networks \cite{tran2023probstar}, so a ProbStar whose predicate is a box is a zonotope with a truncated Gaussian on its coefficients.
Constrained Kalman filters can likewise truncate the Gaussian estimate to a fixed constraint set and re-match its moments \cite{simon2010kalman,simon2010constrained}.
Our Gaussian zonotent instead truncates the Gaussian to the zonotope that bounds the states and propagates both together.

\subsubsection*{Probabilistic Reachable Sets}
A probabilistic reachable set contains the state at one time with at least a prescribed probability, so guarantees over a horizon rely on union bounds.
Stochastic model predictive control bounds the error state by probabilistic reachable sets, such as confidence ellipsoids of its covariance \cite{hewing2018stochastic}.
For nonlinear stochastic systems, a noise-free reachable set can be dilated by a closed-form bound on the stochastic deviation \cite{jafarpour2025probabilistic}.
Stochastic reachability computes the probability of reaching or avoiding a set by dynamic programming over the full transition kernel \cite{abate2008probabilistic}, which scales poorly with the state dimension.
For linear systems with additive noise, bounded or not, forward stochastic reachability gives the exact density of the state through characteristic functions \cite{vinod2017forward}.
Unlike the exact forward density, a zonotent approximates each convolution by matching its mean and covariance or bounds the convolution by a scaled Gaussian, but keeps a support whose order can be reduced.
Reductions of a Gaussian overbound give probabilistic reachable sets for the true state (\Cref{prop: first exit}).

\subsubsection*{Risk over Reachable Sets}
Risk-aware planners keep a system's reachable set robust and evaluate the probability that an uncertain avoid set intersects it \cite{liu2023radius}, or bound the risk of a Gaussian state with chance constraints \cite{blackmore2011chance}.
In these methods, the system's reachable set is a domain of integration with no density of its own, whereas a zonotent's density on it gives the probability that the system violates a constraint.

\section{Preliminaries}\label{sec: preliminaries}

We now present notation and common zonotope operations that our zonotent representation builds on.

\subsubsection*{Notation}
The $n$-dimensional real numbers are $\R^n$, and the nonnegative reals are $\R_{\geq 0}$.
Scalars, vectors, matrices, and sets are italic, random variables are upright sans-serif, e.g., $\rv{x}$, with italic realizations, e.g., $x$, and set-valued functions and probability distributions are uppercase calligraphic, e.g., the zonotope $\zonotopefn(\zcenter,\zgen)$ and the Gaussian distribution $\normaldist(\gaussmean,\gausscov)$ defined below.
Random events are sets of outcomes, so they are also italic, e.g., the event $E$ with indicator $\mathbbm{1}_E$.
An upright superscript label gives context to a scalar, e.g., $\nred$ is the number of generators that probabilistic order reduction keeps.
We write $(v_1,\ldots,v_n)$ for the concatenation of column vectors.
The absolute value $\abs{\cdot}$ and the inequalities $\leq$ and $\geq$ act elementwise on vectors and matrices.
The vector of ones is $\ones$, the vector or matrix of zeros is $\zeros$, and the identity matrix is $\eye$.
The norm $\norm{\cdot}$ without a subscript is Euclidean, the volume (Lebesgue measure) of a set is $\vol(\cdot)$, and probability and expectation are $\Pr(\cdot)$ and $\expect[\cdot]$.

\subsubsection*{Zonotopes}
A zonotope $\zonotope \subset \R^n$ with center $\zcenter \in \R^n$ and generator matrix $\zgen \in \R^{n\times m}$ is an affine map of a hypercube,
\begin{align}
    \zonotope = \zonotopefn(\zcenter,\zgen) = \{\zcenter + \zgen\zcoef \given \norm{\zcoef}_\infty \leq 1\}.
\end{align}
We write $\zgen = [\zgenvec_1,\ldots,\zgenvec_m]$, and for an index set $S \subseteq \{1,\ldots,m\}$, the matrix $\zgen_S$ collects the generators $\zgenvec_k$ with $k \in S$.
The \textit{order} of $\zonotope$ is $m/n$, its number of generators per dimension.
A zonotope with $n$ linearly independent generators is a \textit{parallelotope}.
Zonotopes are closed under affine maps and Minkowski sums:
\begin{align}
    \affmat\zonotope + \affvec &= \zonotopefn(\affmat\zcenter + \affvec, \affmat\zgen), \\
    \zonotope_1 \minksum \zonotope_2 &= \zonotopefn(\zcenter_1 + \zcenter_2, [\zgen_1, \zgen_2]).
\end{align}
\section{Proposed Method: Zonotents}\label{sec: method zonotents}

We now define our proposed zonotent representation, its affine map and Minkowski sum, which are the two operations of set propagation, and its intersection with a set, which collision and goal checks use.

\subsection{Definition}\label{subsec: zonotent definition}
\begin{definition}\label{def: zonotent}
    A \defemph{zonotent} is a zonotope $\zonotope \subset \R^n$ paired with a probability density function (PDF) $\pdf: \R^n \to \R_{\geq 0}$, such that the zonotope is the support of the distribution (i.e., $\int_{\zonotope} \pdf(x)\dx = 1$), which we write as $\zonotent(\zonotope,\pdf)$.
\end{definition}

Two of our operations, namely inner order reduction (\Cref{sec: order reduction}) and set propagation bounding the density of the true state from above (\Cref{sec: first exit}), change the total density mass to not be 1.
Hence, the following definition.

\begin{definition}\label{def: relaxed zonotent}
    A \defemph{relaxed zonotent} $\relaxedzonotent(\zonotope,\pdf)$ pairs a zonotope $\zonotope \subset \R^n$ with a function $\pdf: \R^n \to \R_{\geq 0}$ whose \defemph{mass} $\int_{\zonotope} \pdf(x)\dx$ need not be $1$ but is positive.
\end{definition}

\subsection{Affine Map}\label{sec: affine map}
An affine map moves the support of a zonotent as a zonotope and transforms its PDF by the change of variables.
For every zonotent $\zonotent(\zonotope,\pdf)$ with $\zonotope \subset \R^n$, an invertible $\affmat \in \R^{n\times n}$, and $\affvec \in \R^n$,
\begin{align}\begin{split}\label{eq: affine map}
    (\affmat,\affvec)\cdot\zonotent(\zonotope,\pdf) = \zonotent*(\affmat\zonotope + \affvec,
            x \mapsto \frac{1}{\abs{\det{\affmat}}}\pdf(\affmat\inv(x-\affvec))
        ).
\end{split}\end{align}
The change of variables preserves integrals, so the map keeps the mass of a relaxed zonotent.
\Cref{subsec: gaussian affine map} gives the image of a Gaussian zonotent.

\subsection{Minkowski Sum}\label{subsec: zonotent minkowski sum}
The sum of two independent random variables with zonotent distributions is supported on the Minkowski sum of their supports and has the convolution of their PDFs.

\begin{definition}\label{def: minkowski sum}
    The \defemph{Minkowski sum} of two zonotents or relaxed zonotents $\zonotent(\zonotope_1,\pdf_1)$ and $\zonotent(\zonotope_2,\pdf_2)$ in $\R^n$ is $\zonotent(\zonotope_1,\pdf_1) \minksum \zonotent(\zonotope_2,\pdf_2) = \zonotent(\zonotope_1 \minksum \zonotope_2, \pdf_1 \conv \pdf_2)$, where the convolution of the PDFs is
    \begin{align}\label{eq: convolution}
        (\pdf_1 \conv \pdf_2)(x) = \int_{\zonotope_1 \cap (x - \zonotope_2)} \pdf_1(y)\pdf_2(x-y)\,dy
    \end{align}
    for $x \in \zonotope_1 \minksum \zonotope_2$, and zero elsewhere.
\end{definition}
\noindent If $\pdf_i$ has mass $\zmass_i$, then $\pdf_1 \conv \pdf_2$ has mass $\zmass_1\zmass_2$.

\subsection{Intersection with a Set}\label{subsec: zonotent intersection}\label{sec: intersections}
Checking constraints on zonotent reachable sets of a dynamical system requires computing the mass of the intersection of a zonotent with a set; for example, integrating a PDF over an avoid set or goal set in reach-avoid control.
Where robust methods forbid/require overlap with avoid/goal sets, zonotents additionally enable bounding the probability.

\begin{definition}\label{def: zonotent intersection}
    The \defemph{mass of the intersection} of a zonotent or relaxed zonotent $\zonotent(\zonotope,\pdf)$
    with a measurable set $U \subseteq \R^n$ is
    \begin{align}\label{eq: intersection mass}
        \intmass = \int_{\zonotope \cap U} \pdf(x)\dx.
    \end{align}
\end{definition}
\noindent
For a zonotent, $\intmass = \Pr(\rv{x} \in U)$ for $\rv{x}$ with PDF $\pdf$, which is the probability that the state lands in $U$ (e.g., a goal or avoid set).
The reductions of \Cref{sec: order reduction} preserve these probabilities to within their bounds, which hold for every measurable set.
For a Gaussian zonotent (\Cref{sec: gaussian zonotents}) in $\R^2$ and a polygonal avoid set, $\intmass$ is an integral over a polygon in closed form (\Cref{subsec: planar gaussian mass}).

\section{Gaussian Zonotents}\label{sec: gaussian zonotents}\label{subsec: gaussian zonotents}
We now make zonotents practical by instantiating a specific type of zonotent using a Gaussian PDF supported on a zonotope $\zonotope = \zonotopefn(\zcenter,\zgen) \subset \R^n$ with $\zgen \in \R^{n\times m}$.
We give the affine map of a Gaussian zonotent, the Gaussian mass of a zonotope (for normalizing) and of a polygon, and sampling; we treat the Minkowski sum later in \Cref{sec: minkowski sum}.

\begin{definition}\label{def: gaussian zonotent}
    A \defemph{Gaussian zonotent} has a scaled Gaussian PDF atop a uniform wall, parameterized by a wall share $\wallshare \in [0,1]$, a mean $\gaussmean \in \R^n$ (which need not lie in $\zonotope$), and a positive definite covariance matrix $\gausscov \in \R^{n\times n}$.
    With the PDF $\normaldist(x;\gaussmean,\gausscov)$ of the Gaussian distribution $\normaldist(\gaussmean,\gausscov)$ of mean $\gaussmean$ and covariance $\gausscov$, the PDF of the Gaussian zonotent is
    \begin{align}\label{eq: gaussian zonotent pdf}
        \pdf(x) = \wall + \gaussscale\,\normaldist(x;\gaussmean,\gausscov),
    \end{align}
    where the wall height $\wall = \wallshare/\vol(\zonotope)$ gives the wall the share $\wallshare$ of the mass, and the scale $\gaussscale$ gives the Gaussian the remaining share,
    \begin{align}\label{eq: gaussian scale}
        \gaussscale = \frac{1 - \wallshare}{\gaussmass{\zonotope}},\quad
        \gaussmass{\zonotope} = \int_{\zonotope} \normaldist(x;\gaussmean,\gausscov)\dx.
    \end{align}
\end{definition}

\begin{definition}\label{def: uniform zonotent}
    A \defemph{uniform zonotent} is a Gaussian zonotent with wall share $\wallshare = 1$, for which the scale $\gaussscale$ is zero and the PDF is uniform, $\pdf(x) = 1/\vol(\zonotope)$ on $\zonotope$, whatever the mean and covariance.
\end{definition}

A uniform zonotent models a quantity known only by its bounds (assumed to be a zonotope).
At the other end, with $\wallshare = 0$, the PDF is a truncated and normalized Gaussian.

\begin{remark}\label{rem: wall share default}
    The parameters $\gaussmean$ and $\gausscov$ are not the mean and covariance of the PDF, because of the wall and the truncation to $\zonotope$, and Gaussian zonotents with different wall shares can have the same mean and covariance.
\end{remark}

\subsubsection*{Mode of a Gaussian Zonotent}
For a wall share below $1$, assuming $\gausscov$ invertible, the PDF peaks at the point of $\zonotope$ nearest the mean in the Mahalanobis distance,
\begin{align}\label{eq: gaussian mode}
    \densitymode = \argmin_{x \in \zonotope}\ (x-\gaussmean)\trans\gausscov\inv(x-\gaussmean),
\end{align}
a convex quadratic program in $\zcoef$ with $x = \zcenter + \zgen\zcoef$ and $\norm{\zcoef}_\infty \leq 1$, whose optimal point $\densitymode$ is unique and is the \textit{mode}, where $\pdf$ takes its peak $\densitypeak = \wall + \gaussscale\,\normaldist(\densitymode;\gaussmean,\gausscov)$.

\subsection{Affine Map of a Gaussian Zonotent}\label{subsec: gaussian affine map}
Under the affine map of \Cref{eq: affine map}, the image of a Gaussian zonotent is the Gaussian zonotent with the wall height $\wall/\abs{\det\affmat}$, the mean $\affmat\gaussmean + \affvec$, and the covariance $\affmat\gausscov\affmat\trans$.
Since $\vol(\affmat\zonotope + \affvec) = \abs{\det\affmat}\vol(\zonotope)$, the map keeps the wall share $\wallshare$, so the image of a uniform zonotent is uniform.
The Gaussian mass $\gaussmass{\affmat\zonotope + \affvec}$ under the mapped Gaussian equals $\gaussmass{\zonotope}$, so the scale $\gaussscale$ is unchanged.

\subsection{Gaussian Mass of a Zonotope}\label{subsec: gaussian mass}
Computing the scale $\gaussscale$ requires the Gaussian mass $\gaussmass{\zonotope}$, so we propose strategies to compute or approximate it.

\subsubsection{Parallelotope Splitting}
Every zonotope splits into parallelotopes with disjoint interiors, one for each index set $S$ in the collection $\mathcal{S}$ of the $n$-element index sets whose generators are linearly independent \cite[Sec.~5]{shephard1974combinatorial}
(which underlies the volume formula $\vol(\zonotope) = 2^n\sum_{S \in \mathcal{S}}\abs{\det \zgen_S}$), and each is a box after an affine change of variables, so \Cref{prop: gaussian mass} turns $\gaussmass{\zonotope}$ into a sum of Gaussian probabilities of boxes.

\begin{figure}[t]
    \centering
    \resultfigure{\columnwidth}{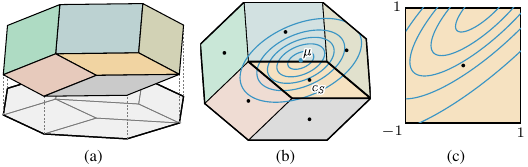}
    \caption{
    Parallelotope split of a zonotope, which gives its Gaussian mass (\Cref{prop: gaussian mass}).
    (a) The lifted zonotope in $\R^3$, whose lower facets (colored) project onto $\zonotope$ (gray).
    (b) The parallelotopes $P_S$ that tile $\zonotope$, with their centers (dots), the center $\zcenter_S$ of the outlined one, and level curves of $\normaldist(\gaussmean, \gausscov)$ (blue).
    (c) The outlined parallelotope in its own coordinates $\zgen_S\inv(x - \zcenter_S)$, where it is the box $[-1,1]^2$.}
    \label{fig: parallelotope split}
    \vspace*{-0.3cm}
\end{figure}

Splitting lifts each generator $\zgenvec_k$ by a height $\liftheight_k$ into $\R^{n+1}$ and projects the lower facets of the lifted zonotope back onto $\zonotope$ (\Cref{fig: parallelotope split}), and heights in general position make every such facet a parallelotope.
Just as the generator matrix $\zgen_S$ collects columns of $\zgen$ (\Cref{sec: preliminaries}), $\liftheight_S$ collects the entries of the heights $\liftheight \in \R^m$ with indices in $S$.

\begin{proposition}\label{prop: gaussian mass}
    Let $\zonotope = \zonotopefn(\zcenter,\zgen) \subset \R^n$, where $\zgen = [\zgenvec_1,\ldots,\zgenvec_m] \in \R^{n\times m}$ has rank $n$, and let $\rv{x} \sim \normaldist(\gaussmean,\gausscov)$.
    Choose heights $\liftheight \in \R^m$ such that, for every index set $T$ with more elements than $\rank(\zgen_T)$, no $a \in \R^n$ satisfies $\zgen_T\trans a = \liftheight_T$.
    Almost every $\liftheight \in \R^m$ qualifies.
    Let $\mathcal{S}$ contain the $n$-element index sets $S$ with $\det \zgen_S \neq 0$, and for each $S \in \mathcal{S}$, let $a_S = \zgen_S\invtrans\liftheight_S$ and $\zcenter_S = \zcenter - \sum_{k \notin S} \sign(\liftheight_k - \zgenvec_k\trans a_S) \zgenvec_k$.
    Then the parallelotopes $P_S = \zonotopefn(\zcenter_S, \zgen_S)$, for $S \in \mathcal{S}$, cover $\zonotope$ and have pairwise disjoint interiors, and
    \begin{align}\label{eq: gaussian mass}
        \gaussmass{\zonotope} = \Pr(\rv{x} \in \zonotope) = \sum_{S \in \mathcal{S}} \Pr\left(\rv{y}_S \in [-1,1]^n\right),
    \end{align}
    where $\rv{y}_S \sim \normaldist\big(\zgen_S\inv(\gaussmean - \zcenter_S),\, \zgen_S\inv\gausscov \zgen_S\invtrans\big)$.
\end{proposition}
\begin{proof}
    For $x \in \zonotope$, the lowest lifted point above $x$ solves the linear program (LP) that minimizes $\liftheight\trans\zcoef$ over $\zcoef \in [-1,1]^m$ with $\zcenter + \zgen\zcoef = x$, and complementary slackness at a vertex $a_S$ of its dual optimal set places $x$ in $P_S$.
    In the interior of $P_S$, the dual optimum is $a_S$ alone, so the $P_S$ cover $\zonotope$ with pairwise disjoint interiors.
    Their boundaries have zero volume, so $\Pr(\rv{x} \in \zonotope) = \sum_{S \in \mathcal{S}} \Pr(\rv{x} \in P_S)$, and $\rv{x} \in P_S$ iff the Gaussian $\rv{y}_S = \zgen_S\inv(\rv{x} - \zcenter_S)$ lies in $[-1,1]^n$.
\end{proof}

Evaluating \Cref{eq: gaussian mass} takes at most $\binom{m}{n}$ box probabilities, which we compute by numerical integration \cite{genz1992numerical}.
Since this grows combinatorially with the number of generators, order reduction (\Cref{sec: order reduction}) is needed to keep the Gaussian mass tractable during set propagation.

\subsubsection{Gaussian Mass of a Polygon}\label{subsec: planar gaussian mass}\label{sec: planar gaussian mass}\label{sec: planar}
In $\R^2$, which covers many uses of zonotents (e.g., position of a mobile robot or angle and velocity of a linear liquid slosh model in \Cref{subsec: results slosh}), the support $\zonotope$, its intersection with a convex polygonal avoid set (\Cref{subsec: zonotent intersection}), and its part in a halfspace are convex polygons, which we compute exactly by clipping \cite{sutherland1974reentrant} from the at most $2m$ vertices of $\zonotope$, given by sorting its generators by angle in $O(m\log m)$ time \cite{michaux2024redefined,jordan2022zonotope}.
The mass of a Gaussian zonotent on a convex polygon $P \subseteq \zonotope$ is $\wall\vol(P) + \gaussscale\,\gaussmass{P}$, with the area $\vol(P)$ from the shoelace formula and the Gaussian mass $\gaussmass{P}$ in closed form in Owen's $T$ function \cite{owen1956tables}, which our implementation uses.

\subsubsection{Halfspace Bounds}
In higher than 2-D, we can bound $\gaussmass{\zonotope}$ from below by Gaussian masses of halfspaces, which have a closed form in every dimension.
For $\rv{x} \sim \normaldist(\gaussmean,\gausscov)$ and a halfspace with normal $\hsnormal \in \R^n$ and offset $\hsoffset$, the scalar $\hsnormal\trans\rv{x}$ is Gaussian with mean $\hsnormal\trans\gaussmean$ and variance $\hsnormal\trans\gausscov\hsnormal$, so
\begin{align}\label{eq: halfspace mass}
    \Pr(\hsnormal\trans\rv{x} \leq \hsoffset) = \stdnormcdf\left(\frac{\hsoffset - \hsnormal\trans\gaussmean}{\sqrt{\hsnormal\trans\gausscov\hsnormal}}\right),
\end{align}
where $\stdnormcdf$ is the standard normal cumulative distribution function.
The halfspace representation of $\zonotope$ writes it as the intersection of the halfspaces $\hsnormal_j\trans x \leq \hsoffset_j$, $j = 1,\ldots,\nfacets$, which we call its \textit{rows}.
A point lies outside $\zonotope$ iff it violates a row, so the union bound over the rows gives
\begin{align}\label{eq: halfspace mass bound}
    \gaussmass{\zonotope} \geq 1 - \sum_{j=1}^{\nfacets}\stdnormcdf\left(-\frac{\hsoffset_j - \hsnormal_j\trans\gaussmean}{\sqrt{\hsnormal_j\trans\gausscov\hsnormal_j}}\right).
\end{align}
Order reduction (\Cref{sec: order reduction}) and the moment-matched sum (\Cref{sec: minkowski sum}) use this guaranteed lower bound.
It is tight when the facets are few and far from the mean in standard deviations, and it loosens as their number grows, up to $2\binom{m}{n-1}$ for $m$ generators.

A second lower bound needs no halfspace representation.
With $u_k\trans$ the rows of $\zgen\trans(\zgen\zgen\trans)\inv$, the coefficient box $\{x \given \abs{u_k\trans(x - \zcenter)} \leq 1,\ k = 1,\ldots,m\}$ lies in $\zonotope$, since each of its points is $\zcenter + \zgen\zcoef$ with $\zcoef_k = u_k\trans(x - \zcenter)$, and \Cref{eq: halfspace mass} on its $2m$ faces bounds $\gaussmass{\zonotope}$ from below at a cost of order $mn^2$.

\subsection{Sampling}\label{sec: sampling}

Since zonotents represent distributions, it is reasonable to ask how one might sample from such a distribution.
The Monte Carlo ground truth in our experiments samples the initial states and the noise from zonotents (\Cref{subsec: results collision,subsec: results narrow gap}) by rejection sampling, which is exact.
For each sample, the sampler picks the wall (i.e., the uniform part) with probability $\wallshare$ and otherwise the truncated Gaussian.
The wall takes candidates uniformly from the axis-aligned box $\{x \given \abs{x - \zcenter} \leq \abs{\zgen}\ones\}$, which contains $\zonotope$, while the truncated Gaussian takes them from $\normaldist(\gaussmean,\gausscov)$, and either keeps the first fresh candidate that lands in $\zonotope$.
When $\gaussmass{\zonotope}$ is small, the truncated Gaussian instead keeps each uniform candidate $x \in \zonotope$ from the box with probability $\normaldist(x;\gaussmean,\gausscov)/\normaldist(\densitymode;\gaussmean,\gausscov)$, with the mode $\densitymode$ of \Cref{eq: gaussian mode}.

\section{Minkowski Sums of Gaussian Zonotents}\label{sec: minkowski sum}

The Minkowski sum of two Gaussian zonotents is supported on a zonotope, but its PDF, the convolution of the two PDFs, is neither uniform nor Gaussian.
To make Gaussian zonotent propagation closed (i.e., return a Gaussian zonotent), we approximate the convolution by a \textit{moment-matched sum} (\Cref{subsec: minkowski moment matched}), which has the convolution's mass and moments, but discards a guarantee that probabilities are over- or under-approximated.
A density pointwise above the convolution would give such a guarantee, but its mass exceeds one, and would be multiplied again by numbers greater than 1 during reachability analysis set propagation, so \Cref{sec: first exit} instead bounds the probability of first exit, that is, that a system's true state ever leaves a sequence of sets.

In this section, we first define our moment-matched sum, then provide ways to compute it, and finally provide a small validation in the plane.
Throughout this section, we write $\convpdf = \pdf_1 \conv \pdf_2$ and $\zonotope = \zonotope_1 \minksum \zonotope_2 = \zonotopefn(\zcenter,\zgen)$ with $\zcenter = \zcenter_1 + \zcenter_2$ and $\zgen = [\zgen_1,\zgen_2]$, $\densitypeak_i$ for the peak of $\pdf_i$, and $\zmass_i$ for the mass of $\pdf_i$ (\Cref{subsec: zonotent minkowski sum}), which is one unless the input is a relaxed zonotent.

\subsection{Moment-Matched Sum}\label{subsec: minkowski moment matched}
The mean and covariance of the convolution are known exactly, so we approximate $\convpdf$ by the Gaussian zonotent on $\zonotope$ that has the same mean and covariance.
Let $\momentmean_i$ and $\momentcov_i$ be the mean and covariance of the normalized PDF $\pdf_i/\zmass_i$, which differ from the parameters $\gaussmean_i$ and $\gausscov_i$ of a Gaussian zonotent (\Cref{rem: wall share default}).

\begin{lemma}\label{lem: convolution moments}
    The normalized convolution $\convpdf/(\zmass_1\zmass_2)$ has mean $\momentmean_1 + \momentmean_2$ and covariance $\momentcov_1 + \momentcov_2$.
\end{lemma}
\begin{proof}
    The normalized convolution is the PDF of $\rv{x} + \rv{y}$ for independent $\rv{x}$ and $\rv{y}$ with PDFs $\pdf_1/\zmass_1$ and $\pdf_2/\zmass_2$.
    Means add, and covariances add because $\rv{x}$ and $\rv{y}$ are independent.
\end{proof}

Next, we explain how to compute the moments of a zonotent, then define the moment-matched sum.

\subsubsection*{Moments of a Zonotent}
We can compute $\momentmean_i$ and $\momentcov_i$ on the parallelotopes $P_S = \zonotopefn(\zcenter_S, \zgen_S)$, $S \in \mathcal{S}$, of \Cref{prop: gaussian mass} (dropping the index $i$ for ease of notation).
Note that the uniform PDF on $P_S$ has mean $\zcenter_S$ and covariance $\zgen_S\zgen_S\trans/3$, since its generator coefficients are independent and uniform on $[-1,1]$, with variance $1/3$.
The uniform PDF on $\zonotope$, which is the wall of a Gaussian zonotent, is the mixture of these with the volume shares $\tileshare_S = 2^n\abs{\det\zgen_S}/\vol(\zonotope)$, and its mean is the center of $\zonotope$ by central symmetry, so
\begin{align}\label{eq: uniform moments}\begin{split}
    \momentmean &= \zcenter,\\
    \momentcov &= \sum_{S \in \mathcal{S}}\tileshare_S\left(\zcenter_S\zcenter_S\trans + \tfrac{1}{3}\zgen_S\zgen_S\trans\right) - \momentmean\momentmean\trans,
\end{split}\end{align}
which is exact and takes at most $\binom{m}{n}$ terms.
By \Cref{eq: gaussian scale}, the PDF \Cref{eq: gaussian zonotent pdf} of a Gaussian zonotent is the mixture of this uniform PDF, with weight $\wallshare$, and of $\normaldist(\gaussmean,\gausscov)$ truncated to $\zonotope$ and normalized, with weight $1 - \wallshare$, so its moments are the same mixture of their moments.
The moments of the truncated Gaussian are integrals of $\normaldist(x;\gaussmean,\gausscov)$ times $1$, $x$, and $xx\trans$ over $\zonotope$.
When $\zonotope$ has $n$ generators whose coefficients are independent under the Gaussian, such as for noise $\normaldist(\zeros,\zgen\zgen\trans/9)$ truncated to $\zonotopefn(\zeros,\zgen)$, they are the exact moments of a product of one-dimensional truncated normals.
The moment-matched sum under mild truncation of \Cref{subsec: computing moment matched} uses these exact moments, where \textit{mild truncation} means that each support cuts off little of its Gaussian, i.e., $\gaussmass{\zonotope}$ is close to one.
Otherwise, each is a sum over the $P_S$ of an integral over a box after the change of variables of \Cref{prop: gaussian mass}, which we compute by a tensor Gauss--Legendre rule with $q$ nodes per axis, whose $\binom{m}{n}q^n$ evaluations, with $q = 16$, limit it to low dimensions.
We now define the sum:

\begin{definition}\label{def: moment matched sum}
    The \defemph{moment-matched sum} of two Gaussian zonotents, possibly relaxed, is the Gaussian zonotent on $\zonotope = \zonotope_1 \minksum \zonotope_2$ of mass $\zmass_1\zmass_2$ and wall share $\wallshare = 0$ with mean $\momentmean_1 + \momentmean_2$ and covariance $\momentcov_1 + \momentcov_2$.
\end{definition}

\subsection{Computing the Moment-Matched Sum}\label{subsec: computing moment matched}

We now provide a fixed point iteration and a closed-form approximation to compute the moment-matched sum parameters $\momentmean$ and $\momentcov$.

\subsubsection*{Fixed-Point Matching}
The moment-matched sum has the mean and covariance of $\convpdf$ by \Cref{lem: convolution moments}, and per \Cref{rem: wall share default} it needs no wall, regardless of the wall shares of its inputs.
So a propagated zonotent is a truncated Gaussian from its first sum on.
Because of the truncation, its parameters $\gaussmean$ and $\gausscov$ differ from its moments (\Cref{rem: wall share default}), so we find them by a fixed-point iteration.
Starting from the targets, each iteration step computes the moments $\momentmean$ and $\momentcov$ of the current Gaussian zonotent as above and adds the residuals $\momentmean_1 + \momentmean_2 - \momentmean$ to $\gaussmean$ and $\momentcov_1 + \momentcov_2 - \momentcov$ to $\gausscov$, halving the step until $\gausscov$ stays positive definite.
We stop at a small residual, at a step limit, or when no match exists (\Cref{rem: moment matching existence}).

\subsubsection*{Moment-Matched Sum under Mild Truncation}
Our moment computation above limits our fixed-point iteration to low dimensions; and, it is not needed when a zonotent's support cuts off little of the Gaussian.
Thus we use a \defemph{moment-matched sum under mild truncation} to set $\gaussmean$ and $\gausscov$ to the target mean and covariance directly.
We form the targets in the same way, taking the moments of each Gaussian input without a wall to be its parameters $\gaussmean_i$ and $\gausscov_i$, while a uniform input and a product of truncated normals as above keep their exact moments.

The error of the sum under mild truncation is that of a Gaussian truncated to a convex set, which we bound by the Gaussian mass outside the set.
We use a classical upper bound on the covariance \cite{kanter1977reduction,brascamp1976extensions} and make explicit the constants in known bounds, of order $\outsidemass\sqrt{\log(1/\outsidemass)}$ and $\outsidemass\log(1/\outsidemass)$, on how much a Gaussian's mean and covariance change when it is conditioned on an event of probability at least $1 - \outsidemass$ \cite[Sec.~2.2.1, after Prop.~2.3]{diakonikolas2023algorithmic}.

\begin{lemma}\label{lem: truncation bound}
    Let $\rv{x} \sim \normaldist(\gaussmean,\gausscov)$ with $\gausscov = \covroot\covroot\trans$, let $\zonotope$ be convex with $\Pr(\rv{x} \notin \zonotope) \leq \outsidemass < 1$, and let $\momentmean$ and $\momentcov$ be the mean and covariance of $\rv{x}$ conditioned on $\rv{x} \in \zonotope$.
    Then
    \begin{align}\label{eq: truncation bound}
        \norm{\covroot\inv(\momentmean - \gaussmean)} \leq \truncmeanerr,\quad
        (1 - \trunccoverr)\gausscov \preceq \momentcov \preceq \gausscov,
    \end{align}
    with $\truncmeanerr = \stdnormcdf'(z_1)/(1 - \outsidemass)$ and $\trunccoverr = \outsidemass + 2z_2\stdnormcdf'(z_2) + \truncmeanerr^2$, where $\stdnormcdf'$ is the standard normal PDF, $z_1 = \stdnormcdf\inv(1 - \min\{\outsidemass, \tfrac{1}{2}\})$, and $z_2 = \stdnormcdf\inv(1 - \outsidemass/2)$.
\end{lemma}
\begin{proof}
    We reduce both bounds to a scalar standard normal along each direction, where, among the events of probability at least $1 - \outsidemass$, conditioning on the complement of one tail moves the mean the most and conditioning on the complement of both tails shrinks the variance the most \cite[Sec.~2.2.1]{diakonikolas2023algorithmic}.
    The upper bound $\momentcov \preceq \gausscov$ holds for a Gaussian conditioned on any convex set \cite{kanter1977reduction}, \cite[Thm.~5.1]{brascamp1976extensions}.
\end{proof}

For the result of the sum, $\gaussmean$ and $\gausscov$ are the targets and $\momentmean$ and $\momentcov$ its own moments, so \Cref{eq: truncation bound} bounds their distance.
Because the targets take each Gaussian input's moments to be its parameters, \Cref{lem: truncation bound} applied to each input also bounds the error in the targets.
We take $\outsidemass$ from the larger of the bound of \Cref{eq: halfspace mass bound} and the coefficient-box bound of \Cref{subsec: gaussian mass}, and our implementation reports $\outsidemass$, $\truncmeanerr$, and $\trunccoverr$ with every such sum.

\begin{remark}\label{rem: moment matching existence}
    A Gaussian zonotent with the target moments need not exist, since truncating Gaussians to $\zonotope$ does not give every covariance of a PDF on $\zonotope$, and the iteration then stops at its step limit with a large $\gausscov$.
    We do not prove that the iteration converges when a match exists.
    See numerical results in \Cref{subsec: results minkowski}.
\end{remark}

\subsection{Validating Moment-Matched Sum Accuracy in $\R^2$}\label{subsec: results minkowski}
We check the moment-matched sum against the exact convolution $\convpdf$ of \Cref{eq: convolution} on five pairs of zonotents in $\R^2$ with three and two generators, three Gaussian (weakly truncated, strongly truncated, and with walls) and two with a uniform first input.
The $L^1$ distance of the sum to $\convpdf$ was $0.025$, $0.09$, and $0.08$ for the Gaussian pairs and $0.12$ and $0.37$ for the pairs with a uniform input, after $2$ to $13$ iterations and at most \qty{5}{\ms}.
The sum is thus close to the convolution wherever the convolution is close to a Gaussian, and farthest when a large uniform support and a narrow Gaussian give the convolution a flat top, which no Gaussian follows.

\section{Probabilistic Order Reduction}\label{sec: order reduction}

\begin{figure}[t]
    \centering
    \resultfigure{\columnwidth}{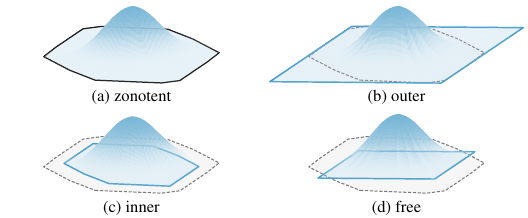}
    \caption{Probabilistic order reduction of one Gaussian zonotent in $\R^2$, with the density drawn above its support.
    (a) The zonotent, with support $\zonotope$ (black outline).
    (b) The outer reduction encloses $\zonotope$ (dashed) in a reduced support $\redzono$ (blue outline).
    (c) The inner reduction lies in $\zonotope$ and discards the part outside $\redzono$ (gray).
    (d) The free reduction is neither inside nor around $\zonotope$.}
    \label{fig: reduction overview}
\end{figure}

We now propose \textit{probabilistic order reduction} to leverage a zonotent's density to simplify its support zonotope $\zonotope$.
Our goal is to replace $\zonotope$, which has $m$ generators, to a zonotope $\redzono$ with $\nred$ generators, where $n \leq \nred \leq m$ (\Cref{fig: reduction overview}).
Each method keeps different guarantees on bounding reachable states and giving the probability of any subset.
First, outer reduction encloses the support, $\zonotope \subseteq \redzono$, as in order reduction for zonotopes \cite{girard2005reachability,kopetzki2017methods}.
Second, inner reduction lies in the support, $\redzono \subseteq \zonotope$, as in \cite{lutzow2025underapproximative}, and uses the density to discard a part of the support that carries at most a budget $\discardmass \in (0,1)$ of the probability.
Third, free reduction drops both containments and only keeps $1 - \discardmass$ of the probability.
\Cref{fig: reduction overview} uses the budget $\discardmass = 0.05$, far larger than in the experiments, so that the discarded parts are visible.
We state all three for a Gaussian zonotent without a wall, noting we discard the wall by default in a moment-matched sum (\Cref{def: moment matched sum}).

To proceed, we first define the covariance-aligned parallelotopes that the outer and free reductions use, then state our three reduction strategies.

\subsection{Covariance-Aligned Parallelotopes}\label{subsec: aligned parallelotopes}

Our outer and free reductions use parallelotopes whose generators are aligned with the covariance for ease of numerical implementation, since the Gaussian mass of such a parallelotope is a product of one-dimensional masses.
Let $\covroot$ be a square root of the covariance, so that $\gausscov = \covroot\covroot\trans$, and let $\covrot \in \R^{n\times n}$ be a rotation, so that $\covroot\covrot$ is also a square root.
If $\rv{x}$ is sampled from $\normaldist(\gaussmean,\gausscov)$, then $\rv{y} = (\covroot\covrot)\inv(\rv{x} - \gaussmean)$ is standard normal, and we call $y = (\covroot\covrot)\inv(x - \gaussmean)$ the whitened coordinates.
A parallelotope with the generator matrix $\covroot\covrot\diag(\halfwidth)$, for half-widths $\halfwidth \in \R^n$, is a box in these coordinates, so its Gaussian mass is a single box probability with independent coordinates.

We make frequent use of two such parallelotopes.
The first is the enclosing parallelotope, which is the interval hull of $\zonotope$ in whitened coordinates,
\begin{align}\label{eq: enclosing parallelotope}
    \redzono = \zonotopefn\big(\zcenter,\, \covroot\covrot\diag(\halfwidth)\big),\quad
    \halfwidth_j = \sum_{k=1}^{m}\abs{\basisvec_j\trans(\covroot\covrot)\inv\zgenvec_k},
\end{align}
where $\basisvec_j$ is the $j$th unit vector and $\zgenvec_1,\ldots,\zgenvec_m$ are the generators of $\zonotope$.
The second is the confidence parallelotope, the cube of radius $\confradius > 0$ about the mean in whitened coordinates,
\begin{align}\label{eq: confidence parallelotope}
    \redzono_{\confradius,\covrot} = \zonotopefn(\gaussmean,\, \confradius\covroot\covrot),
\end{align}
whose Gaussian mass and volume are
\begin{align}\label{eq: confidence parallelotope mass}
    \gaussmass{\redzono_{\confradius,\covrot}} = \big(2\stdnormcdf(\confradius) - 1\big)^n,\quad
    \vol(\redzono_{\confradius,\covrot}) = (2\confradius)^n\sqrt{\det\gausscov},
\end{align}
for every rotation $\covrot$, with $\stdnormcdf$ as in \Cref{eq: halfspace mass}.
The radius thus sets both the mass and the volume, and the Gaussian mass is $1 - \discardmass$ at the radius
\begin{align}\label{eq: confidence radius}
    \confradius_{\discardmass} = \stdnormcdf\inv\Big(\tfrac{1}{2} + \tfrac{1}{2}(1 - \discardmass)^{1/n}\Big).
\end{align}
We choose $\covroot\covrot$ among six square roots of the covariance (the Cholesky factor, the symmetric one, the one along the principal axes, and three rotations of the Cholesky factor computed from the whitened generators) and keep the one whose enclosing parallelotope has the least volume.

\subsection{Outer Reduction}\label{subsec: outer reduction}
An outer reduction replaces the support by a zonotope $\redzono \supseteq \zonotope$ with $\nred$ generators and keeps the parameters of the density, so the reduced zonotent is the Gaussian zonotent on $\redzono$ with the same mean and covariance.
No probability is discarded, and every state in $\zonotope$ is in $\redzono$, so every robust statement made with the support still holds.
Any outer order reduction for zonotopes gives $\redzono$ \cite{girard2005reachability,kopetzki2017methods}.
We also use the enclosing parallelotope of \Cref{eq: enclosing parallelotope}, with $n$ generators.

Enlarging the support renormalizes the density over a larger set, which changes each probability by at most the mass that moves to the added part.
For a Gaussian zonotent without a wall, the reduced density is $\gaussmass{\zonotope}/\gaussmass{\redzono}$ times the original on $\zonotope$ and puts the rest of its mass, $1 - \gaussmass{\zonotope}/\gaussmass{\redzono}$, on $\redzono\setminus\zonotope$.
It therefore changes the probability of any set by at most $1 - \gaussmass{\zonotope}$, the Gaussian mass outside the original support, however large $\redzono$ is.
The lower bound of \Cref{eq: halfspace mass bound} on $\gaussmass{\zonotope}$ bounds this change without computing a Gaussian mass of a zonotope.
The added volume does make the robust statements more conservative, and it accumulates over repeated reductions as for zonotopes.

\subsection{Inner Reduction}\label{subsec: inner reduction}
An inner reduction discards a part of the support that holds little probability, which gives a smaller support in which every state is still reachable but which no longer contains every reachable state.
Given the budget $\discardmass$, we seek a zonotope $\redzono \subseteq \zonotope$ with $\nred$ generators that holds at least $1 - \discardmass$ of the mass of $\pdf$.
Inner reduction thus returns the restriction $\redpdf = \pdf$ on $\redzono$, a relaxed zonotent whose mass is the kept mass $\zmass' = \int_{\redzono}\pdf\dx$, which is below the mass $\zmass$ of the original.
Note, in numerical implementation, inner reduction may not meet the budget.

To proceed, we first bound the change in probability resulting from inner reduction.
We then instantiate inner reduction using a \textit{template zonotope} that fits inside the original support; we propose two constructions of $\redzono$ on it, the volume template and the mass template.

\subsubsection{Probability Bounds}
The reduced density can either keep $\pdf$ on $\redzono$ or be rescaled to the original mass, and both bound probabilities computed with the original zonotent:

\begin{proposition}\label{prop: reduction guarantee}
    Let $\zonotent(\zonotope, \pdf)$ be a zonotent with mass $\zmass = \int_{\zonotope}\pdf\dx$, which is $1$ before any reduction, and let $\redzono \subseteq \zonotope$ have the kept mass $\zmass' = \int_{\redzono}\pdf\dx \geq (1 - \discardmass)\zmass$.
    If $\redpdf = \pdf$ on $\redzono$, then, for every measurable set $U$,
    \begin{align}\label{eq: reduction guarantee}
        \int_{U \cap \redzono}\redpdf\dx \leq \int_{U}\pdf\dx \leq \int_{U \cap \redzono}\redpdf\dx + \discardmass\zmass.
    \end{align}
    If instead $\redpdf = (\zmass/\zmass')\pdf$ on $\redzono$, which rescales the reduced zonotent to the mass $\zmass$, then, for every measurable set $U$,
    \begin{align}\label{eq: reduction guarantee rescaled}
        \abs{\int_{U \cap \redzono}\redpdf\dx - \int_{U}\pdf\dx} \leq \zmass - \zmass' \leq \discardmass\zmass.
    \end{align}
\end{proposition}
\begin{proof}
    Write $\int_U\pdf\dx = a + b$, with $a = \int_{U \cap \redzono}\pdf\dx \in [0, \zmass']$ and $b = \int_{U \setminus \redzono}\pdf\dx$, which is nonnegative and at most $\int_{\zonotope \setminus \redzono}\pdf\dx = \zmass - \zmass' \leq \discardmass\zmass$.
    If $\redpdf = \pdf$, the integral of $\redpdf$ over $U \cap \redzono$ is $a$, which gives \Cref{eq: reduction guarantee}.
    If $\redpdf$ is rescaled, that integral is $(\zmass/\zmass')a$, and its difference from $a + b$ is $(\zmass/\zmass' - 1)a - b$, which lies between $-b \geq -(\zmass - \zmass')$ and $(\zmass/\zmass' - 1)\zmass' = \zmass - \zmass'$.
\end{proof}

Any $\redzono \subseteq \zonotope$ satisfies the hypothesis with $\discardmass$ replaced by the share $1 - \zmass'/\zmass$ that it discards.
So, \Cref{prop: reduction guarantee} holds with that share even for a reduction that misses its budget.
Below, $\discardmass_k$ denotes the share discarded (at most its budget, if met by the reduction) by the $k$th reduction of a zonotent during reachability propagation.

The guarantee holds through exact set propagation, because exact affine maps and convolutions with a PDF do not increase the $L^1$ distance between two densities.
By the triangle inequality, after $K$ reductions, rescaled or not, a probability is thus within $\sum_k\discardmass_k$ of the one computed without reductions.
These bounds are relative to the zonotent being reduced, not to the true distribution, and do not hold through the moment-matched sum of \Cref{def: moment matched sum}, which only approximates the convolution.

\subsubsection{Template Zonotope}
We make a template zonotope by fixing the directions of the reduced support's generators but leaving their lengths free, so making $\redzono$ becomes a choice of $\nred$ lengths and a center.
We take the $\nred$ largest generators of $\zonotope$, with indices $\tmplidx$, skipping near-parallel ones and, until $n$ are chosen, any that does not raise the rank.
Note \cite{lutzow2025underapproximative} previously proposed fixing the directions and choosing the lengths under a linear containment condition; zonotents enable doing this scaling by mass.
The template is $\redzono_{\shapescale} = \zonotopefn(\tmplcenter, \tmplgen\diag(\shapescale))$, where the columns of $\tmplgen$, the largest generators, are the $\zgenvec_k$ for $k \in \tmplidx$, the scales $\shapescale > 0$ set their lengths, and $\tmplcenter$ is its center.
We write $\redzono_{\sizescale\shapescale}$ for the template $\redzono_{\shapescale}$ shrunk about its center by a size $\sizescale \in (0,1]$.

The template must lie in the original support.
For each row $\hsnormal_j\trans x \leq \hsoffset_j$, $j = 1,\ldots,\nfacets$, of the halfspace representation of $\zonotope$, the largest value of $\hsnormal_j\trans x$ over the template is $\hsnormal_j\trans\tmplcenter + \sum_{k\in\tmplidx}\shapescale_k\abs{\hsnormal_j\trans\zgenvec_k}$, so the template lies in $\zonotope$ iff
\begin{align}\label{eq: template containment}
    \containmat\shapescale \leq \containvec,\quad \containmat_{jk} = \abs{\hsnormal_j\trans\zgenvec_k},\quad \containvec_j = \hsoffset_j - \hsnormal_j\trans\tmplcenter.
\end{align}
This condition is linear in $\shapescale$ and $\tmplcenter$ together.
It has a strictly feasible point whenever $\tmplcenter$ lies in the interior of $\zonotope$.

The template center rule places the center $\tmplcenter$ at the mode $\densitymode$ of \Cref{eq: gaussian mode}, where $\pdf$ is largest, when the mode lies in the interior of $\zonotope$.
Otherwise, the mode is on the boundary of $\zonotope$, where a template has no room, so the template is centered at the center $\zcenter$ of the support.
Next, we provide volume- and mass-based ways to pick the scales $\shapescale$.

\paragraph{Volume Template}
The volume template chooses the shape $\shapescale$ that gives the template the largest volume inside $\zonotope$, and then its size by mass.
For $\nred = n$, the template is a parallelotope of volume $2^n\abs{\det\tmplgen}\prod_k\shapescale_k$, so maximizing its volume is the concave program
\begin{align}\label{eq: template shape}
    \max_{\shapescale > 0}\ \sum_{k\in\tmplidx}\log\shapescale_k \quad\st\ \containmat\shapescale \leq \containvec,
\end{align}
which we solve by a barrier method.
For $\nred > n$, the volume is a sum of such products over the $n$-element subsets of $\tmplidx$; we keep \Cref{eq: template shape} as a concave surrogate.

The volume-maximizing shape ignores $\pdf$, so it may hold far more than $1 - \discardmass$ of the mass.
We shrink it about $\tmplcenter$ to $\redzono_{\sizescale\shapescale}$ and seek the smallest size that keeps the budget,
\begin{align}\label{eq: template size}
    \min_{\sizescale \in (0,1]}\ \sizescale \quad\st\ \int_{\redzono_{\sizescale\shapescale}}\pdf\dx \geq (1 - \discardmass)\int_{\zonotope}\pdf\dx.
\end{align}
The sets $\redzono_{\sizescale\shapescale}$ are nested, with volume $\sizescale^n\vol(\redzono_{\shapescale})$, so the kept mass is nondecreasing in $\sizescale$, and bisection finds the member of least volume that holds $1 - \discardmass$ of the mass.
If even the template $\redzono_{\shapescale}$ at $\sizescale = 1$ holds less than $1 - \discardmass$ of the mass, \Cref{eq: template size} is infeasible, and we keep $\sizescale = 1$.
The kept mass of $\redzono_{\sizescale\shapescale}$ is $\zmass\gaussscale\gaussmass{\redzono_{\sizescale\shapescale}}$, by \Cref{prop: gaussian mass} on only $\binom{\nred}{n}$ parallelotopes.
The volume template is the cheapest inner reduction, since it needs one concave program and one bisection, but it may fail to produce a result:

\begin{remark}\label{rem: reduction volume}
    Since its shape ignores $\pdf$, a volume template whose directions are poorly aligned with the mass may fail to hold $1 - \discardmass$ even at $\sizescale = 1$.
\end{remark}

\paragraph{Mass Template}
The mass template keeps the directions and the center of the template and chooses its scales by mass.
It avoids the failure of \Cref{rem: reduction volume} whenever some scales of the template meet the budget.
We seek the scales that keep the most mass,
\begin{align}\label{eq: scales max mass}
    \max_{\shapescale > 0}\ \int_{\redzono_{\shapescale}}\pdf\dx \quad\st\ \containmat\shapescale \leq \containvec,
\end{align}
whose optimal value says whether the template can meet the budget.
That value is at least the mass of the volume-maximizing shape of \Cref{eq: template shape}, which is feasible.
The set $\redzono_{\shapescale}$ is linear in $\shapescale$ under Minkowski sums, so, by the Pr\'ekopa--Leindler inequality \cite[Thm.~7.1]{gardner2002brunn}, which implies that a measure with a log-concave density is log-concave \cite[Thm.~2]{prekopa1973logarithmic}, the kept mass is log-concave in $\shapescale$, and \Cref{eq: scales max mass} is a convex program.
If the optimal value meets the budget, we shrink the solution about its center to the budget, as in \Cref{eq: template size}.

We solve \Cref{eq: scales max mass} numerically and reduce the returned scales if needed so that \Cref{eq: template containment} holds, so the solver affects only the volume and the mass of the reduction, not its guarantee.

\begin{remark}\label{rem: mass shape limits}
    An inner reduction needs room in the support for a zonotope that holds $1 - \discardmass$ of the mass.
    The confidence parallelotope that holds $1 - \discardmass$ of the Gaussian mass has the radius $\confradius_{\discardmass}$ of \Cref{eq: confidence radius}, which grows with the dimension $n$, while the largest cube that fits in the support in whitened coordinates need not.
    The budget can thus be out of reach of an inner reduction in higher dimensions, or when the support is small relative to the covariance, as for a zonotent that is close to uniform.
\end{remark}

\subsection{Free Reduction}\label{subsec: free reduction}
A free reduction asks only for $\nred$ generators and the budget, and requires neither $\redzono \subseteq \zonotope$ nor $\zonotope \subseteq \redzono$, so its support has the potential to leave out reachable states and contain unreachable ones.
We take the confidence parallelotope $\redzono = \redzono_{\confradius_{\discardmass},\covrot}$ of \Cref{eq: confidence parallelotope} at the radius of \Cref{eq: confidence radius}, with the same mean and covariance, whose Gaussian mass is $1 - \discardmass$ by \Cref{eq: confidence parallelotope mass}.
The parallelotope exists in closed form in every dimension, and rotation changes neither its mass nor its volume, so we use the square root of \Cref{subsec: aligned parallelotopes} to align it with the support.

Whether or not its output fits in the support, free reduction changes any probability by at most the Gaussian mass that the two supports leave out:

\begin{proposition}\label{prop: free reduction}
    Let $\pdf$ and $\redpdf$ be the PDF of $\normaldist(\gaussmean,\gausscov)$ restricted to $\zonotope$ and to $\redzono$ and normalized.
    Then, for every measurable set $U$,
    \begin{align}\label{eq: free reduction}
        \abs{\int_{U\cap\redzono}\redpdf\dx - \int_{U\cap\zonotope}\pdf\dx} \leq \big(1 - \gaussmass{\zonotope}\big) + \big(1 - \gaussmass{\redzono}\big).
    \end{align}
\end{proposition}
\begin{proof}
    Restricting $\normaldist(\gaussmean,\gausscov)$ to $\zonotope$ and normalizing changes the probability of $U$ by $\big(1/\gaussmass{\zonotope} - 1\big)\gaussmass{U\cap\zonotope} - \gaussmass{U\setminus\zonotope}$, a difference of two terms in $[0, 1 - \gaussmass{\zonotope}]$.
    The same holds for $\redzono$, and the triangle inequality gives the claim.
\end{proof}
\noindent 
Note that the bound does not depend on where $\redzono$ lies relative to $\zonotope$, so it holds even when the two supports barely overlap.
For the confidence parallelotope, the second term is the budget $\discardmass$, and the first is bounded by \Cref{eq: halfspace mass bound}.
Through exact operations, a chain of free reductions (i.e., during reachable set propagation) changes a probability by at most the sum of the bounds of \Cref{eq: free reduction}, not the sum of the budgets $\discardmass_k$, since each bound also counts the Gaussian mass outside the support being reduced.

Compared to inner reduction, free reduction is useful because, when the support is symmetric about the mean, we can meet the budget in every dimension, without the Gaussian mass of the support:

\begin{proposition}\label{prop: free reduction share}
    Let $\zonotope$ and $\redzono$ be zonotopes with center $\gaussmean$, and let $\pdf$ be the PDF of $\normaldist(\gaussmean,\gausscov)$ restricted to $\zonotope$ and normalized.
    Then $\int_{\zonotope\cap\redzono}\pdf\dx \geq \gaussmass{\redzono}$, which is $1 - \discardmass$ for the confidence parallelotope of radius $\confradius_{\discardmass}$.
\end{proposition}
\begin{proof}
    Both sets are convex and symmetric about $\gaussmean$, so $\gaussmass{\zonotope\cap\redzono} \geq \gaussmass{\zonotope}\,\gaussmass{\redzono}$ by the Gaussian correlation inequality \cite[(1.1)]{royen2014simple}, and dividing by $\gaussmass{\zonotope}$ gives the claim.
\end{proof}

The support is symmetric about the mean whenever the initial set and the noise are symmetric about their means, since affine maps and Minkowski sums preserve this.
Otherwise, since $\gaussmass{\zonotope\cap\redzono} \geq \gaussmass{\zonotope} - \big(1 - \gaussmass{\redzono}\big)$, we ask the confidence parallelotope for the smaller budget $\discardmass\ul{\nu}$, where $\ul{\nu} \leq \gaussmass{\zonotope}$ is $\gaussmass{\zonotope}$ in $\R^2$ and the bound of \Cref{eq: halfspace mass bound} otherwise, which guarantees the share $1 - \discardmass$ whenever $\ul{\nu} > 0$.

\section{First-Exit Bounds}\label{sec: first exit}

We now bound the probability that a system's true state ever leaves a sequence of sets constructed during reachability analysis; since probabilistic order reduction could discard parts of the reachable set, this notion of \textit{first exit} gives us a diagnostic tool to check if indeed the system entered the discarded part.
Our proof is for uncertain affine systems; we leave extension to nonlinear systems (which are well-studied with zonotope reachability analysis \cite{althoff2021set}) to future work.

Our idea is to propagate a function that lies above the density, instead of a zonotent that approximates it (i.e., we use a relaxed zonotent).
Roughly speaking, if each propagated function lies above the density of the true state, then the mass it puts outside a set is at least the probability that the true state first leaves the sets there.

The following proposition is the main result, since its bound holds for the true system rather than for the propagated zonotent.
We then give such functions in closed form, the Gaussian overbound, and apply the bound with two kinds of sets, the reduced supports of \Cref{sec: order reduction} and the states in which a system is safe.
In the proposition, $\mappedpdf_t$ is the image of $\overpdf_t\mathbbm{1}_{\keepset_t}$ under $x \mapsto \affmat_t x + \affvec_t$, mapped as in \Cref{eq: affine map}.

\begin{proposition}\label{prop: first exit}
    Let $\rv{x}_{t+1} = \affmat_t\rv{x}_t + \affvec_t + \noisegain_t\rv{\procnoise}_t$ for $t = 0, 1, \ldots$, with each $\affmat_t \in \R^{n \times n}$ invertible and each $\noisegain_t \in \R^{n \times m}$ of any rank, where $\rv{x}_0$ and the independent $\rv{\procnoise}_t$ have the PDFs $\truepdf_0$ on $\R^n$ and $\truepdf_{\rv{\procnoise}_t}$ on $\R^m$, and let $\keepset_0, \keepset_1, \ldots \subseteq \R^n$ be measurable sets (e.g., zonotent supports).
    Let $\overpdf_0 \geq \truepdf_0$ and $\overpdf_{\rv{\procnoise}_t} \geq \truepdf_{\rv{\procnoise}_t}$ pointwise, and let each $\overpdf_{t+1}$ satisfy
    \begin{align}\label{eq: first exit recursion}
        \overpdf_{t+1}(x) \geq \int_{\R^m} \mappedpdf_t(x - \noisegain_t y)\,\overpdf_{\rv{\procnoise}_t}(y)\,dy \quad \forall\, x \in \R^n.
    \end{align}
    Then, for every number of steps $\nsteps$,
    \begin{align}\label{eq: first exit}
        \Pr\big(\rv{x}_t \notin \keepset_t \text{ for some } t \leq \nsteps\big) \leq \sum_{t=0}^{\nsteps}\int_{\R^n \setminus \keepset_t} \overpdf_t\dx.
    \end{align}
\end{proposition}
\begin{proof}
    Let $\insidepdf_t$ be the density of $\rv{x}_t$ on the event that $\rv{x}_s \in \keepset_s$ for every $s < t$, so that $\insidepdf_0 = \truepdf_0$.
    As $\rv{\procnoise}_t$ is independent of $\rv{x}_0, \ldots, \rv{x}_t$, the density $\insidepdf_{t+1}$ is the right side of \Cref{eq: first exit recursion} with $\insidepdf_t$ and $\truepdf_{\rv{\procnoise}_t}$ in place of $\overpdf_t$ and $\overpdf_{\rv{\procnoise}_t}$, which is a density on $\R^n$ even when $\noisegain_t$ has rank less than $n$, since $\affmat_t$ is invertible.
    That right side is nondecreasing in both functions, since its integrand is a product of nonnegative functions, so $\insidepdf_t \leq \overpdf_t$ by induction.
    The state first leaves the sets at step $t$ with probability $\int_{\R^n\setminus\keepset_t} \insidepdf_t\dx$, and summing over $t$ gives the claim.
\end{proof}
\noindent Note, the sets need not lie in the supports, and the functions $\overpdf_t$ need not have unit mass, since only their masses outside the sets enter \Cref{eq: first exit}.

We now provide a \textit{Gaussian overbound} to give such functions in closed form when $\rv{x}_0$ and each $\rv{\procnoise}_t$ are Gaussians truncated to zonotopes $\zonotope_0$ and $\zonotope_{\rv{\procnoise}_t}$.
The PDF of such a term is at most $1/\gaussmass{\cdot}$ times its untruncated Gaussian, where $\gaussmass{\cdot}$ is the mass of its zonotope under that Gaussian, so we take $\overpdf_{\rv{\procnoise}_t}$ to be this multiple of the Gaussian on $\zonotope_{\rv{\procnoise}_t}$.
Let $\gaussscale_0 = 1/\gaussmass{\zonotope_0}$ and $\gaussscale_{t+1} = \gaussscale_t/\gaussmass{\zonotope_{\rv{\procnoise}_t}}$, and let $\gaussmean_{t+1} = \affmat_t\gaussmean_t + \affvec_t + \noisegain_t\gaussmean_{\rv{\procnoise}_t}$ and $\gausscov_{t+1} = \affmat_t\gausscov_t\affmat_t\trans + \noisegain_t\gausscov_{\rv{\procnoise}_t}\noisegain_t\trans$ be the mean and covariance of the untruncated terms propagated through the system.
Then the relaxed Gaussian zonotent $\overpdf_t = \gaussscale_t\normaldist(\cdot\,;\gaussmean_t,\gausscov_t)$ on $\zonotope_t$ satisfies \Cref{eq: first exit recursion} for any zonotopes $\zonotope_{t+1} \supseteq \affmat_t(\zonotope_t \cap \keepset_t) \minksum (\affvec_t + \noisegain_t\zonotope_{\rv{\procnoise}_t})$, since a convolution of Gaussians is Gaussian, also when $\noisegain_t$ has rank less than $n$.
We take $\zonotope_{t+1} = \affmat_t\zonotope_t \minksum (\affvec_t + \noisegain_t\zonotope_{\rv{\procnoise}_t})$, the exact reachable zonotope, or $\zonotope_{t+1} = \affmat_t\keepset_t \minksum (\affvec_t + \noisegain_t\zonotope_{\rv{\procnoise}_t})$ when $\keepset_t$ is a zonotope.
This overbound loosens only through the scale $\gaussscale_t$, which grows by the factor $1/\gaussmass{\zonotope_{\rv{\procnoise}_t}}$ at each step.
For example, for the truncated Gaussian noise of the narrow gap of \Cref{subsec: results narrow gap}, $\gaussscale_{12} = 1.007$, so the overbound loosens by under one percent over twelve steps.

With the reduced supports as the sets, the bound is the total mass discarded by the reductions.
An inner or free reduction of $\overpdf_t$ to $\keepset_t = \redzono_t$ that discards $\exitmass_t = \int_{\zonotope_t \setminus \redzono_t} \overpdf_t\dx$ makes the right side of \Cref{eq: first exit} equal to $\sum_t \exitmass_t$, since $\overpdf_t$ vanishes outside $\zonotope_t$, with no moment matching.
The reduced overbound is not rescaled to unit mass, since \Cref{eq: first exit recursion} needs the restriction of $\overpdf_t$ to the reduced support.

With the states in which a system is safe as the sets $\keepset_t$, the Gaussian overbound drops the restriction to $\keepset_t$ in \Cref{eq: first exit recursion}, since a Gaussian restricted to a set is no longer Gaussian.
Integrating it over the whole of $\zonotope_t \setminus \keepset_t$ at every step then amounts to a union bound over the steps, which counts a trajectory once for every step that it spends outside the sets, rather than only at the step where it first leaves them.
Instead, at each step, we can integrate the overbound only over the states that one step reaches from the previous set, so that a trajectory that has left the sets is counted again only while it stays within one step of them:

\begin{corollary}\label{cor: first passage}
    Let $\overpdf_t$ be the Gaussian overbound propagated with $\keepset_t = \R^n$, on the exact reachable zonotopes $\zonotope_t$, and let $\reachset_t = \affmat_{t-1}\keepset_{t-1} \minksum (\affvec_{t-1} + \noisegain_{t-1}\zonotope_{\rv{\procnoise}_{t-1}})$ be the set that one step reaches from $\keepset_{t-1}$.
    Then, for every number of steps $\nsteps$,
    \begin{align}\label{eq: first passage}\begin{split}
        \Pr\big(\rv{x}_t \notin \keepset_t \text{ for some } t \leq \nsteps\big) \leq {}&\int_{\zonotope_0 \setminus \keepset_0}\overpdf_0\dx + {}\\
        &\sum_{t=1}^{\nsteps}\int_{(\zonotope_t \cap \reachset_t) \setminus \keepset_t} \overpdf_t\dx.
    \end{split}\end{align}
\end{corollary}
\begin{proof}
    Let $\passpdf_0 = \overpdf_0$, and for $t \geq 1$ let $\passpdf_t$ be the right side of \Cref{eq: first exit recursion} formed from $\overpdf_{t-1}$ and $\keepset_{t-1}$.
    It is at most $\overpdf_t$, which bounds the same integral without the indicator of $\keepset_{t-1}$, and it vanishes outside $\zonotope_t \cap \reachset_t$, since $\overpdf_{\rv{\procnoise}_{t-1}}$ vanishes outside $\zonotope_{\rv{\procnoise}_{t-1}}$.
    Since $\passpdf_t \leq \overpdf_t$, the right side of \Cref{eq: first exit recursion} formed from $\passpdf_t$ is at most the one formed from $\overpdf_t$, which is $\passpdf_{t+1}$, so the functions $\passpdf_t$ satisfy \Cref{prop: first exit}, which then gives the claim.
\end{proof}

Finally, the sets can also be tightened to account for an unknown parameter of the system, such as the fill depth of a sliding cup of water in \Cref{subsec: results slosh}.
Suppose that we propagate the overbound for a nominal model, with the parameter at a fixed value, and that a zonotope $\mismatchset_t$ contains the difference between the state of the true system and the state of the nominal model at step $t$, under the same initial state and noise.
Whenever the nominal state lies in the tightened set $\{x \given x \minksum \mismatchset_t \subseteq \keepset_t\}$, the true state then lies in $\keepset_t$.
So the bound of \Cref{prop: first exit} for the nominal model with the tightened sets also bounds the probability that the true state ever leaves the sets $\keepset_t$, whatever the value of the parameter.
Because the unknown parameter multiplies the state, bounding $\mismatchset_t$ needs a bound on the propagated state, which the support $\zonotope_t$ gives, whereas a Gaussian does not (\Cref{subsec: results slosh}).

\section{Numerical Results}\label{sec: experiments}

We now test zonotents numerically.
We seek to answer five questions:
\begin{enumerate}[label=Q\arabic*]
    \item How do outer, inner, and free reductions behave as system dimension increases?
    Does each reduction meet its budget, and what does each cost in volume, accuracy, and time (\Cref{subsec: results reduction dimension})?

    \item When all noise is bounded, how do the probabilities of unsafe sets of a Gaussian zonotent compare with those of probabilistic zonotopes \cite{althoff2009safety} (\Cref{subsec: results probzono})?
    \item For an affine dynamical system with bounded noise, how close to the truth are the probabilities of collision of a Gaussian zonotent, and how much smaller is its reduced support than the exact reachable zonotope (\Cref{subsec: results collision})?
    
    \item Does the first-exit bound of \Cref{sec: first exit}, with the reduced supports as the sets, certify for the true system a narrower gap than the exact reachable zonotope (\Cref{subsec: results narrow gap})?
    
    \item Is the first-passage bound of \Cref{cor: first passage} accurate on a difficult model?
    When sliding an open cup of water at varying fill depth, does our zonotent model accept faster slides than worst-case reachability without spilling, and does its slosh model agree with a fluid simulation (\Cref{subsec: results slosh})?
\end{enumerate}
\noindent The validation and the experiments use our PyTorch implementation\footnote{See project page: \url{https://zonotents.github.io/}.} on a laptop CPU.
Each Monte Carlo estimate of a probability has a $99\%$ Clopper--Pearson interval \cite{clopper1934use} from its count, per step, so even an exact model misses about one interval in a hundred.

\subsection{Validation: Order Reduction across Dimension}\label{subsec: results reduction dimension}
We address Q1 by validating the outer, inner, and free reductions of \Cref{sec: order reduction} along propagated states from $\R^2$ to $\R^{10}$, to check which meets its budget and what each costs in volume, accuracy, and time.

\subsubsection{Validation Criterion}
A reduction passes if it meets its budget at every step.

\subsubsection{Test Design}
For each $n \in \{2, 3, 4, 6, 8, 10\}$, we sample a linear system $\rv{x}_{t+1} = \affmat\rv{x}_t + \rv{\procnoise}_t$ with a random $\affmat$ of spectral radius $0.9$, an initial state uniform on a box, and noise on a parallelotope $\zonotopefn(\zeros, \zgen_{\rv{\procnoise}})$ that is either uniform or the Gaussian $\normaldist(\zeros, \zgen_{\rv{\procnoise}}\zgen_{\rv{\procnoise}}\trans/9)$ truncated to it, so the exact support of $\rv{x}_t$ has $n(t+1)$ generators.
We propagate each system for $20$ steps by the moment-matched sum under mild truncation, first without reduction, timing each sum and the Gaussian mass of \Cref{prop: gaussian mass} on supports with $n$ to $8n$ generators, and then reducing the support after every step by one method with $\discardmass = 10^{-3}$ and rescaling to unit mass.
The outer reductions are the enclosing parallelotope of \Cref{eq: enclosing parallelotope} and the reduction by principal component analysis (PCA) \cite{kopetzki2017methods}, which encloses the support in a parallelotope along the principal axes of its generators.
The inner reductions are the volume template and the mass template, both with $2n$ generators.
The free reduction is the confidence parallelotope of \Cref{subsec: free reduction}, with $n$ generators.
Every support in these systems is symmetric about its mean, so \Cref{prop: free reduction share} guarantees the budget of the free reduction.
For the same reason, the mass template stays centered at the mode.

At every step, we record whether the reduction meets its budget and the share of simulated states in the reduced support.
We also record the volume of the reduced support over that of the $99.9\%$ confidence ellipsoid of the exact moments.

\begin{table}[t]
    \centering
    \caption{Order reductions along $20$ steps with truncated Gaussian noise and $\discardmass = 10^{-3}$, by dimension $n$}
    \label{tab: results reduction dimension}
    \footnotesize
    \setlength{\tabcolsep}{3pt}
    \begin{tabular}{lcccccc}
        \toprule
        & $n = 2$ & $3$ & $4$ & $6$ & $8$ & $10$ \\
        \midrule
        \multicolumn{7}{l}{\textit{budget met}} \\
        in.\ vol. & 19/19 & 17/19 & 0/19 & -- & -- & -- \\
        in.\ mass & 19/19 & 19/19 & 18/19 & -- & -- & -- \\
        \midrule
        \multicolumn{7}{l}{\textit{states kept}} \\
        in.\ vol. & 0.9989 & 0.9982 & 0.2311 & -- & -- & -- \\
        in.\ mass & 0.9989 & 0.9990 & 0.9991 & -- & -- & -- \\
        free & 0.9989 & 0.9991 & 0.9991 & 0.9991 & 0.9991 & 0.9991 \\
        \midrule
        \multicolumn{7}{l}{\textit{volume}} \\
        out.\ par. & 19 & 104 & $2{\cdot}10^{3}$ & $6{\cdot}10^{5}$ & $5{\cdot}10^{7}$ & $1{\cdot}10^{12}$ \\
        out.\ PCA & 26 & 332 & $1{\cdot}10^{6}$ & $9{\cdot}10^{7}$ & $6{\cdot}10^{11}$ & $7{\cdot}10^{19}$ \\
        in.\ vol. & 1.03 & 1.33 & 0.86 & -- & -- & -- \\
        in.\ mass & 1.06 & 1.16 & 1.72 & -- & -- & -- \\
        free & 1.10 & 1.32 & 1.68 & 3.07 & 6.27 & 13.9 \\
        \bottomrule
    \end{tabular}
    \tablenote{Methods: outer reductions to the enclosing parallelotope (out.\ par.) and by PCA \cite{kopetzki2017methods} (out.\ PCA), the inner volume and mass templates (in.\ vol., in.\ mass), and the free confidence parallelotope (free).
    Blocks of the table show number of reductions that met the budget, the smallest share of the simulated states in the reduced support, and the volume of the support at the last step over that of the $99.9\%$ confidence ellipsoid.
    The templates keep $2n$ generators, so they reduce only at the $19$ steps at which the support has more.
    The states kept are a share of $10^5$ simulated trajectories of the true system, with a standard error of about $10^{-4}$.
    A dash marks a method that timed out or was too large to attempt.
    }
\end{table}

\subsubsection{Results}
Without reduction, every moment-matched sum under mild truncation took at most \qty{40}{\ms}, on supports with up to $210$ generators in $\R^{10}$.
The Gaussian mass finished within \qty{60}{\s} for up to $8n$ generators in $\R^2$ to $\R^4$, but only for $2n$ in $\R^6$ and $\R^8$ and for $n$ in $\R^{10}$.
With truncated Gaussian noise, the free reduction was the only one that discards mass and met its budget in every dimension of \Cref{tab: results reduction dimension}.
The free reduction also met its budget with uniform noise, in $1$ to \qty{2}{\ms} per reduction.
The volume of its support grew with the dimension, but by many orders of magnitude less than that of the outer reductions.

The volume of the outer reductions was still growing at the last step in every dimension.

Among the inner reductions, the mass template took $0.05$ to \qty{2.8}{\s} per reduction from $\R^2$ to $\R^4$ and timed out from $\R^6$ on.
With uniform noise, every inner reduction missed its budget at every step from $\R^3$ on, and in $\R^2$ the two templates met it at $4$ and $13$ of $19$ steps.

\subsubsection{Discussion}
For a practitioner, the free reduction should be used as default beyond $\R^2$, since it met its budget in every dimension within \qty{2}{\ms}, while the inner reductions, which keep the support inside the reachable set, are worth their cost only in low dimensions and with noise close to a truncated Gaussian.
The moment-matched sum scales to $\R^{10}$, whereas the Gaussian mass of a zonotope does not, which is the reason for the halfspace bounds of \Cref{eq: halfspace mass bound} and for the free reduction, which avoids the Gaussian mass.
Free reduction scales because the volume of its support over that of the confidence ellipsoid depends only on $n$ and $\discardmass$, up to the difference between the covariance of the zonotent and the exact one, at the price of a support that is neither inside nor outside the reachable set.
The outer reductions keep the robust statement, but their volume compounds as for zonotopes.
The inner reductions fail in higher dimensions and with uniform noise because no zonotope with few generators inside the support holds $1 - \discardmass$ of the mass (\Cref{rem: reduction volume,rem: mass shape limits}), and each also shrinks the support that the next must fit inside.

\subsection{Probabilistic Zonotopes with Bounded Noise}\label{subsec: results probzono}
We address Q2 with an example of \cite{althoff2009safety} in which we make all noise bounded, comparing the probabilities of unsafe sets of a Gaussian zonotent with Monte Carlo simulation and with the bounds of probabilistic zonotopes.

\subsubsection{Hypothesis}
We hypothesize that, when the noise is bounded, the Gaussian zonotent gives probabilities of unsafe sets closer to Monte Carlo simulation than the upper bound of a probabilistic zonotope \cite{althoff2009safety}, which represents a random state as the sum of a zonotope, whose points carry no distribution, and a Gaussian of covariance $\pzcov$.

\subsubsection{Experiment Design}
We use the two-dimensional linear system of \cite[Sec.~VII]{althoff2009safety},
\begin{align}\label{eq: results probzono system}
    \dot{\rv{x}} = \ctsmat\rv{x} + \pzinput + \ctsnoisegain\ctsnoise,\quad
    \ctsmat = \left[\begin{smallmatrix} -1 & -4 \\ 4 & -1 \end{smallmatrix}\right],\quad
    \ctsnoisegain = 0.7\eye,
\end{align}
with an input $\pzinput$ in a small box and the white noise $\ctsnoise$, sampled exactly with the time step $0.01$ at the steps $k = 0,\ldots,250$.
The initial state, which we take to be $\normaldist([3, 2]\trans, 0.5^2\eye)$, and the discrete-time noise are either truncated to $\pm 3$ standard deviations per axis or replaced by uniform PDFs, on the same box for the initial state and with the same covariance for the noise.
At every step, we evaluate the probabilities of the unsafe set $\unsafeset = \{x \given x_2 \leq -1.5\}$ and of the tail event $x_2 \leq -2.5$ under the Gaussian zonotent with the exact mean and covariance on the exact support and under the untruncated Gaussian.
The probabilistic zonotope bounds each by the integral over the event of its enclosing hull $\hullpdf(x) = \max_{c} \normaldist(x; c, \pzcov)$, where $c$ ranges over its zonotope, which holds the input and, in the uniform case, either the initial box or the smaller box of what we call the \textit{box hull}, the hull of this form with the least mass that encloses the uniform initial PDF, following the enclosure of non-Gaussian initial sets in \cite[Sec.~VI-D]{althoff2009safety}.
We compare with $10^6$ simulated trajectories, which sample the initial state and the noise per axis and the input uniformly on its box.

\subsubsection{Results}
With truncated Gaussian noise, the Gaussian zonotent lay within the $99\%$ Monte Carlo intervals at all but one of the $502$ pairs of a step and an event, while the untruncated Gaussian gave the same probabilities to within $10^{-12}$.
The bound of the probabilistic zonotope on the probability of $\unsafeset$ exceeded the Monte Carlo estimate by at most $0.017$ and lay above its interval at $122$ of the $251$ steps.
With uniform noise, the Gaussian zonotent was outside the intervals at $80$ of the $251$ steps for $\unsafeset$ and at $69$ for the tail event, with errors of at most $0.007$, and the untruncated Gaussian was outside them at $83$ and $69$ steps.
With the initial box as a set, the bound of the probabilistic zonotope exceeded one at $39$ steps, and with the box hull it was at most $0.89$, against a largest Monte Carlo estimate of $0.38$, and lay above the interval at every step.

\subsubsection{Discussion}
The hypothesis holds, though with truncated Gaussian noise the bound of the probabilistic zonotope is nearly as close, within $0.017$ of the Monte Carlo estimate.
With uniform noise, its bound exceeds the estimate by up to $0.52$ with the box hull, while the errors of the Gaussian zonotent are at most $0.007$.
The hull of a probabilistic zonotope places the peak density of its Gaussian part at every point of its zonotope, so with the initial box in its zonotope, its bound exceeds one.
The box hull avoids this, at the price of a loose bound.
So, against both baselines, the Gaussian zonotent matches or improves on the probabilities and adds a compact support that bounds every reachable state.

\subsection{Collision Probability}\label{subsec: results collision}
We address Q3 with a planar affine system whose state passes an obstacle, comparing the probabilities of collision of a Gaussian zonotent with Monte Carlo simulation.

\subsubsection{Hypothesis}
We hypothesize that the Gaussian zonotent propagated by the moment-matched sum, with an inner reduction of its support at every step, gives probabilities of collision as close to the truth as the untruncated Gaussian, on a support several times smaller than the exact reachable zonotope and its outer order reduction.

\subsubsection{Experiment Design}
We consider the state $\rv{x}_t \in \R^2$ of the discrete-time affine system
\begin{align}\label{eq: results system}
    \rv{x}_{t+1} &= \affmat\rv{x}_t + \affvec + \rv{\procnoise}_t, \\
    \affmat &= 0.85\left[\begin{smallmatrix} \cos\theta & -\sin\theta \\ \sin\theta & \cos\theta \end{smallmatrix}\right],\quad
    \affvec = \left[\begin{smallmatrix} 0.6 \\ 0.3 \end{smallmatrix}\right],\nonumber
\end{align}
with $\theta = 20^\circ$.
The initial state is uniform on $\zonotope_0 = \zonotopefn(\zeros, \zgen_0)$ with $\zgen_0 = \left[\begin{smallmatrix} 0.3 & 0.1 \\ 0 & 0.2 \end{smallmatrix}\right]$.
The disturbances $\rv{\procnoise}_t$ are independent and supported on $\zonotope_{\rv{\procnoise}} = \zonotopefn(\zeros, \zgen_{\rv{\procnoise}})$ with $\zgen_{\rv{\procnoise}} = \left[\begin{smallmatrix} 0.12 & 0 & 0.06 \\ 0 & 0.12 & -0.06 \end{smallmatrix}\right]$.
We run two noise models, in which $\rv{\procnoise}_t$ is either uniform on $\zonotope_{\rv{\procnoise}}$ or the Gaussian $\normaldist(\zeros, \zgen_{\rv{\procnoise}}\zgen_{\rv{\procnoise}}\trans/9)$ truncated to $\zonotope_{\rv{\procnoise}}$, a Gaussian zonotent without a wall.
We compute the probability $\Pr(\rv{x}_t \in \obstacle)$ that the state lies in the obstacle $\obstacle = \zonotopefn([1.6, 1.3]\trans, \left[\begin{smallmatrix} 0.35 & 0.1 \\ 0 & 0.3 \end{smallmatrix}\right])$ for $t = 0,\ldots,5$.

The zonotent propagation alternates the exact affine map of \Cref{sec: affine map} with the moment-matched sum with the zonotent of the disturbance, computed by the fixed-point iteration of \Cref{subsec: minkowski moment matched}, which gives a zonotent $\zonotent(\zonotope_t, \hat{\pdf}_t)$.
We run it without reduction, on the exact support $\zonotope_t$ of $\rv{x}_t$, and with the inner reduction by the volume template after every step, with $\discardmass = 10^{-3}$ and $\nred = 4$.
Before the next step, we rescale the reduced zonotent to unit mass, so the bound of \Cref{eq: reduction guarantee rescaled} applies.
The baselines are the untruncated Gaussian, which has the exact mean and covariance of the state and no support, the exact reachable zonotope $\zonotope_t$ (with $2 + 3t$ generators), and its outer order reduction to four generators after every step \cite{girard2005reachability}, which keeps some generators and encloses the others in a parallelotope.
The two zonotopes report only whether a collision is possible.
We integrate $\hat{\pdf}_t$ over the polygon $\zonotope_t \cap \obstacle$, and the untruncated Gaussian over $\obstacle$, exactly with Owen's $T$ function (\Cref{subsec: planar gaussian mass}).
We compare with $10^6$ simulated trajectories, which sample the initial state and the noise with the samplers of \Cref{sec: sampling}.

\subsubsection{Results}
Without reduction, the largest errors of the moment-matched zonotent over $t = 0,\ldots,5$ were $0.004$ with uniform noise and $0.007$ with truncated Gaussian noise, the same as those of the untruncated Gaussian.
The reduction after every step raised the largest error by up to $0.002$.
With both noise models, neither the moment-matched zonotent nor the untruncated Gaussian lay inside the $99\%$ Monte Carlo interval at any of the steps $t = 2$ to $5$, at which the probability of collision is positive, since $10^6$ samples resolve differences well below their errors of at most $0.007$.
With truncated Gaussian noise, the support reduced after every step had area $0.52$ at $t = 5$, against $2.56$ for the exact reachable zonotope and $4.61$ for its outer reduction.
The reduction met its budget at four of the five steps with truncated Gaussian noise, and only at $t = 5$ with uniform noise.

\subsubsection{Discussion}
The hypothesis holds for the probabilities, and for the size of the support only with truncated Gaussian noise, where the reduced support after five steps is $5$ times smaller than the exact reachable zonotope and $9$ times smaller than its outer reduction.
The zonotent takes its accuracy from the exact mean and covariance, as the untruncated Gaussian does, so its density mainly serves the reduction.
With uniform noise, the reduction missed its budget for the reason of \Cref{rem: mass shape limits}, so it should then run less often, keep more generators, or give up containment in the support, as the free reduction does.

\subsection{Narrow Gap}\label{subsec: results narrow gap}
We address Q4 with a mobile robot that passes a narrow gap, whose tracking error follows the system of Q3.
Q3 computes approximate probabilities of collision by the moment-matched sum, which carries no guarantee for the true system, and checks them against simulation.
Here we instead bound the probability of collision for the true system by propagating the Gaussian overbound of \Cref{sec: first exit} and applying \Cref{prop: first exit} with the reduced supports as the sets, and we ask how narrow a gap these supports certify.

\subsubsection{Hypothesis}
We hypothesize that the reduced supports certify passage through a gap much narrower than the one that the exact reachable zonotope certifies, with a probability of collision at most the sum of the budgets.

\subsubsection{Experiment Design}
\begin{figure}[t]
    \centering
    \resultfigure{\columnwidth}{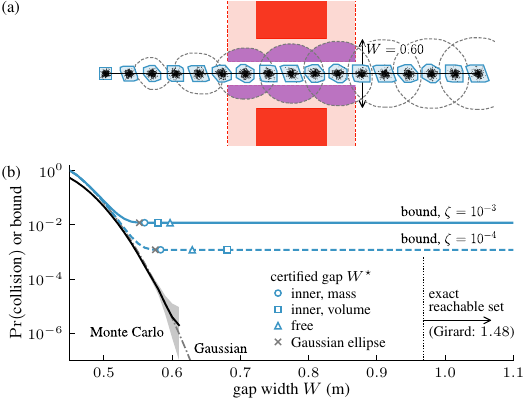}
    \caption{Narrow gap of Q4 with truncated Gaussian noise.
    (a) The obstacles (red) and their enlargement by the body (dashed red), the exact reachable zonotope at every other step (gray dashed) and its overlap with the enlarged obstacles (purple), the supports reduced by the mass template (blue), and simulated positions (dots), for a gap of $W = \qty{0.60}{\m}$.
    (b) Probability of collision against $W$, simulated with its $99\%$ band (black) and under the untruncated Gaussian (gray), the bound of the mass template for two budgets (blue), the gap $W^\star$ that each reduction and the guaranteed ellipses certify (markers), and the narrowest gap that the exact reachable zonotope certifies (dotted).}
    \label{fig: results narrow gap}
\end{figure}

A robot with a rectangular body \qty{0.5}{\m} long and \qty{0.4}{\m} wide and a fixed heading tracks the path $(-1.6 + 0.2t, 0)$, $t = 0,\ldots,16$, between two rectangular obstacles \qty{0.6}{\m} deep that leave a gap of width $W$ (\Cref{fig: results narrow gap}).
Its tracking error follows \Cref{eq: results system} with $\affvec = \zeros$, the noise generators $\zgen_{\rv{\procnoise}}/4$, and truncated Gaussian noise, starting from $\normaldist(\zeros, 0.05^2\eye/9)$ truncated to $\zonotopefn(\zeros, 0.05\eye)$.
A set of positions certifies a gap if the set misses both obstacles enlarged by the body.
For each sequence of sets, such as the reduced supports, we call $W^\star$ the narrowest gap certified by every set in the sequence.
We reduce the support of the Gaussian overbound after every step by the volume and the mass templates with $\nred = 4$ and by the free reduction with $\nred = 2$, each discarding at most $\discardmass \in \{10^{-3}, 10^{-4}\}$ of the mass of $\overpdf_t$, and compute the discarded masses $\exitmass_t$ exactly by clipping and Owen's $T$ function (\Cref{subsec: planar gaussian mass}).
Unlike in Q3, the reduced overbound is not rescaled, so that it still satisfies \Cref{eq: first exit recursion}.
The true state thus leaves a reduced support by step $12$, the last at which the exact reachable set meets the enlarged obstacles, with probability at most $\sum_{t \leq 12} \exitmass_t$, which bounds the probability of collision at every $W \geq W^\star$.
The baselines are the exact reachable zonotope, its outer reduction to four generators after every step, and the confidence ellipses of $\normaldist(\gaussmean_t,\gausscov_t)$ clipped to the exact reachable zonotope, which hold the state with probability at least $1 - \discardmass$ per step because $\overpdf_t$ bounds its PDF.
We check collisions at the time steps against $10^6$ simulated trajectories.

\subsubsection{Results}
With $\discardmass = 10^{-3}$, every reduction met its budget at every step up to $t = 12$, so each guaranteed bound was $0.012$ (\Cref{fig: results narrow gap}).
The volume template, the mass template, and the free reduction certified gaps of $0.58$, $0.56$, and \qty{0.60}{\m}, against \qty{0.97}{\m} for the exact reachable zonotope, \qty{1.49}{\m} for its outer reduction, and \qty{0.55}{\m} for the confidence ellipses.
The narrowest gap with no simulated collision was \qty{0.61}{\m}.
Between $0.76\%$ and $0.79\%$ of the simulated trajectories left the reduced supports by $t = 12$, against the bound of $1.2\%$.
With $\discardmass = 10^{-4}$, the mass template and the ellipses both certified \qty{0.58}{\m} with a bound of $1.2 \times 10^{-3}$.

\subsubsection{Discussion}
The hypothesis holds, and the bound applies to the true system.
The reduced supports certify a gap $42\%$ narrower than the exact reachable zonotope.
The bound on leaving them is within a factor of $1.6$ of the simulated rate, whereas the bound on collision is loose because the bound counts every exit from the reduced supports as a collision.
The guaranteed confidence ellipses certify about the same gap, so in this linear example the density gives the reduction its guarantee rather than a narrower gap.
Unlike the ellipses, the inner reductions keep zonotopes with four generators inside the reachable set, which further zonotope operations can take as input.
Checking collisions only at the time steps leaves out collisions between steps, which a planner would cover by sweeping the body along each step.

\subsection{Sliding a Cup of Water}\label{subsec: results slosh}

Sloshing water is hard to predict, since its free surface moves nonlinearly and its natural frequency depends on how full the cup is.
We address Q5 by choosing how fast to slide an open cup of water across a table without spilling it.
We bound the probability of spilling at every fill depth by \Cref{cor: first passage}, with the states that do not spill as the sets.
As in Q4 and unlike in Q3, this bound holds for the true system, which is here a linear slosh model.
Unlike in Q4, the sets come from the task rather than from order reduction, and the bounded support serves instead to bound the effect of the unknown fill depth.
We compare our bound on a linear slosh model against a three-dimensional fluid simulation.

\subsubsection{Hypothesis}
We hypothesize that the bound on the probability of spilling is at least its Monte Carlo estimate for every plan and fill depth.
We also hypothesize that the bound accepts faster slides than worst-case reachability and a Gaussian tail bound.
Last, we hypothesize that the linear slosh model predicts whether a noise-free minimum-jerk slide brings the water above the rim in a three-dimensional fluid simulation, and when it first does so.

\subsubsection{Experiment Design}
We model the first sloshing mode of water in an upright cylindrical cup of radius $\cupradius$, filled to depth $\filldepth$, by its equivalent linear pendulum \cite{abramson1966dynamic,ibrahim2005liquid}, whose angle $\sloshangle$ obeys
\begin{align}\label{eq: slosh model}
    \ddot{\sloshangle} + 2\sloshdamp\sloshfreq\dot{\sloshangle} + \sloshfreq^2\sloshangle = -\frac{\sloshfreq^2\cupaccel}{\gravity},\quad \sloshfreq^2 = \frac{\gravity\besselroot}{\cupradius}\tanh\frac{\besselroot\filldepth}{\cupradius},
\end{align}
under the horizontal acceleration $\cupaccel$ of the cup, with gravity $\gravity$, the damping ratio $\sloshdamp$, and $\besselroot \approx 1.841$ the first zero of $J_1'$.
The surface at the wall rises $\surfheight = \surfgain\sloshangle$, with $\surfgain = \cupradius$.
Under a zero-order hold of $\samplestep = \qty{10}{\ms}$, $\rv{x}_k = (\sloshangle_k, \dot{\sloshangle}_k)$ evolves as $\rv{x}_{k+1} = \sloshstepmat(\filldepth)\rv{x}_k + \sloshstepinput(\filldepth)(\cupcmd_k + \rv{\acceldist}_k)$, with the commanded acceleration $\cupcmd_k$ and the tracking error $\rv{\acceldist}_k$.
For $\cupradius = \qty{3.5}{\cm}$, a rim \qty{10}{\cm} high, $\sloshdamp = 0.005$, and $\filldepth \in [\filldepthlo, \filldepthhi] = [7.5, 8.5]~\unit{\cm}$, the rim lies $\spillmargin(\filldepth) \in [1.5, 2.5]~\unit{\cm}$ above the water, and $\sloshfreq \approx \qty{22.71}{\radian\per\s}$.
The water spills if $\abs{\surfheight} > \spillmargin$ at a step $k \leq \nsteps$, or if the amplitude $(\sloshangle_{\nsteps}^2 + \dot{\sloshangle}_{\nsteps}^2/\sloshfreq^2)^{1/2}$ at the last step exceeds $\spillmargin/\surfgain$.

The initial angle and rate have standard deviations \qty{0.03}{\radian} and \qty{0.68}{\radian\per\s}, and $\rv{\acceldist}_k$ has $\accelstd_k = \qty{0.1}{\m\per\s\squared} + 0.1\abs{\cupcmd_k}$.
All of these are independent Gaussians truncated to $\pm 3$ standard deviations, while $\filldepth$ is unknown.
A plan slides the cup \qty{40}{\cm} in $\movetime \in \{0.60, 0.61, \ldots, 2.00\}~\unit{\s}$ along a minimum-jerk (MJ) profile, or that profile convolved with a zero-vibration (ZV) shaper \cite{singer1990preshaping,feddema1997control} tuned to the nominal $\sloshfreq$, and then rests for \qty{0.5}{\s}.
Each method picks the shortest plan whose probability of spilling, as it computes it, is at most $\spillrisk \in \{0.01, 0.05\}$.

The bound propagates $\rv{x}_k$ with the matrices $\sloshstepmatref$ and $\sloshstepinputref$ of $\filldepthref = \qty{8}{\cm}$, as the system of \Cref{prop: first exit} with $\affmat_k = \sloshstepmatref$, $\affvec_k = \sloshstepinputref\cupcmd_k$, and the rank-one noise gain $\noisegain_k = \sloshstepinputref$.
Its Gaussian overbound lies on the exact reachable zonotope $\zonotope_k$, with $\gaussscale_k = \truncmass^{-(k+2)}$ for the $k+2$ truncated terms, where $\truncmass = 0.9973$ is the mass within $\pm 3$ standard deviations, the exact mean $\gaussmean_k$, and the covariance $\boundcov_k$ of the untruncated terms.
The sets $\keepset_k$ require $\abs{\sloshangle_k} \leq \spillmargin(\filldepthhi)/\surfgain - \mismatchbound_k$, with a tightening $\mismatchbound_k \geq 0$ given below, and $\keepset_{\nsteps}$ also requires the amplitude to be at most $\spillmargin(\filldepthhi)/\surfgain$.
We integrate the overbound over the sets of \Cref{cor: first passage} in closed form with Owen's $T$ function (\Cref{subsec: planar gaussian mass}).

The fill depth enters through zonotopes $\mismatchset_k$ that contain the difference between the states at $\filldepth$ and at $\filldepthref$ under the same accelerations for every depth in $[\filldepthlo, \filldepthhi]$, as at the end of \Cref{sec: first exit}.
We build them from interval matrices \cite{althoff2007reachability} about $\sloshstepmatref$ and $\sloshstepinputref$, with elementwise radii $\mismatchrad_{\sloshstepmat}$ and $\mismatchrad_{\sloshstepinput}$, over the bounded support $\zonotope_k$, which a Gaussian lacks.
Every entry of $\sloshstepmat(\filldepth)$ and $\sloshstepinput(\filldepth)$ is monotone in $\filldepth$ here, so we take the radii from the two ends of the depth interval.
For a depth $\filldepth \in [\filldepthlo, \filldepthhi]$, let $x_k$ and $x^0_k$ be the states at $\filldepth$ and at $\filldepthref$ from the same initial state under the same accelerations $\cupaccel_k = \cupcmd_k + \acceldist_k$.
Since the noise is truncated to $\pm 3$ standard deviations, $\abs{\acceldist_k} \leq 3\accelstd_k$ and $x^0_k \in \zonotope_k$.
The difference $e_k = x_k - x^0_k$ satisfies $e_0 = \zeros$ and $e_{k+1} = \sloshstepmatref e_k + (\sloshstepmat(\filldepth) - \sloshstepmatref)x_k + (\sloshstepinput(\filldepth) - \sloshstepinputref)\cupaccel_k$.
If $e_k \in \mismatchset_k$, then $\abs{x_k} \leq \maxabs(\zonotope_k \minksum \mismatchset_k)$, where $\maxabs(\set{X})$ is the elementwise largest absolute value over a set $\set{X}$, and $\abs{\cupaccel_k} \leq \abs{\cupcmd_k} + 3\accelstd_k$.
The last two terms of $e_{k+1}$ then lie in the box of the recursion below, so by induction on $k$, $e_k \in \mismatchset_k$ for every $k$, where $\mismatchset_0 = \{\zeros\}$ and
\begin{align}\label{eq: slosh mismatch set}\begin{split}
    \mismatchset_{k+1} = {}&\sloshstepmatref\mismatchset_k \minksum \zonotopefn\big(\zeros, \diag(\mismatchrad_{\sloshstepmat}\maxabs(\zonotope_k \minksum \mismatchset_k) + {}\\
    &\mismatchrad_{\sloshstepinput}(\abs{\cupcmd_k} + 3\accelstd_k))\big).
\end{split}\end{align}

With $\mismatchbound_k = \max_{e \in \mismatchset_k}\abs{e_1}$, a spill at any depth, where $\spillmargin(\filldepth) \geq \spillmargin(\filldepthhi)$, puts the trajectory at $\filldepthref$ with the same noise outside some $\keepset_k$, so the bound holds at every depth.

Monte Carlo takes the largest spill frequency of $10^6$ trajectories at each of the depths $7.5$, $8$, and \qty{8.5}{\cm}.
The Gaussian tail sums the tails of $\normaldist(\gaussmean_k, \boundcov_k)$ over the steps, as chance-constrained planning does \cite{blackmore2011chance}.
Whenever the mean $\gaussmean_k$ lies within the threshold, as it does for every accepted plan, these per-step tails bound the probability at $\filldepthref$, since each truncated term is more peaked than its Gaussian \cite[Thm.~1]{birnbaum1948peakedness}.
As a reference without a guarantee, the Gaussian with the first-passage bound integrates $\normaldist(\gaussmean_k, \boundcov_k)$ over the same sets without clipping them to $\zonotope_k$.
Worst-case reachability \cite{girard2005reachability} requires $\zonotope_k \minksum \mismatchset_k$ to stay within the threshold, whereas the nominal check, as in input shaping, requires only the noise-free slosh at \qty{8}{\cm} to stay below the rim.

To check the slosh model itself, we simulate the water in the cup with Basilisk \cite{popinet2009accurate}.
We apply noise-free MJ commands in $30$ slides on cells of \qty{1.25}{\mm}, most of them near the model's spill boundary, and in $33$ slides on cells of \qty{2.5}{\mm} to check the grid.
In each slide, we compare the height at the wall with the model's $\surfgain\sloshangle$.

\begin{table}[t]
    \centering
    \caption{Shortest accepted slide durations for the cup in Q5}
    \label{tab: results slosh}
    \setlength{\tabcolsep}{2.5pt}
    \begin{tabular}{lcccc}
        \toprule
        & \multicolumn{2}{c}{MJ} & \multicolumn{2}{c}{ZV} \\
        \cmidrule(lr){2-3}\cmidrule(lr){4-5}
        method & $\spillrisk = 0.01$ & $0.05$ & $0.01$ & $0.05$ \\
        \midrule
        Monte Carlo & 1.03 (8.3) & 1.00 (32) & 1.01 (8.7) & 0.98 (33) \\
        zonotent bound$^{\mathrm{a}}$ & 1.05 (3.7) & 1.01 (20) & 1.03 (3.3) & 0.99 (22) \\
        Gaussian, first passage & 1.04 (5.6) & 1.00 (32) & 1.02 (5.5) & 0.98 (33) \\
        Gaussian tail$^{\mathrm{b}}$ & 1.06 (2.6) & 1.02 (13) & 1.04 (2.0) & 1.01 (8.7) \\
        \bottomrule
    \end{tabular}
    \tablenote{Each entry is the duration in seconds on a \qty{0.01}{\s} grid, with its Monte Carlo spill probability in units of $10^{-3}$ in parentheses.
    ZV includes the \qty{0.14}{\s} of the shaper.
    Rows marked $^{\mathrm{a}}$ bound the spill probability at every depth, rows marked $^{\mathrm{b}}$ bound it per step at $\filldepthref$ along these plans, and the Gaussian with the first-passage bound carries no guarantee.}
\end{table}

\begin{figure}[t]
    \centering
    \resultfigure{\columnwidth}{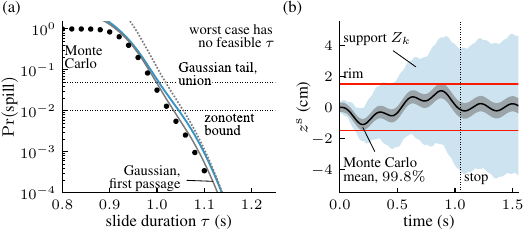}
    \caption{Probability of spilling and slosh height for the cup of Q5.
    (a) Probability of spilling of MJ slides against $\movetime$, from Monte Carlo (dots), the zonotent bound (blue), and the Gaussian baselines (gray), with $\spillrisk = 0.01$ and $0.05$ (dotted).
    (b) Surface height at the wall for the plan that the zonotent bound picks at $\spillrisk = 0.01$, with the support $\zonotope_k$ (blue), the mean and central $99.8\%$ of simulated trajectories (gray), and the rim at $\filldepthhi$ (red).}
    \label{fig: results slosh}
\end{figure}

\begin{figure}[t]
    \centering
    \resultfigure{\columnwidth}{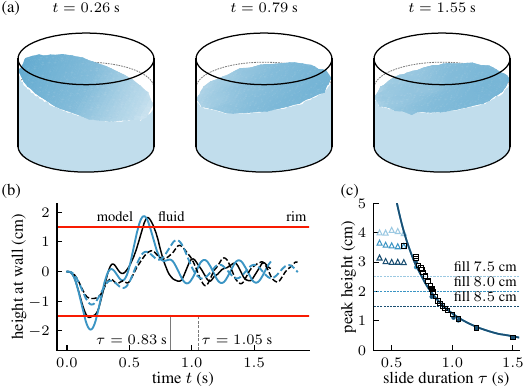}
    \caption{Fluid simulation of noise-free MJ slides in Basilisk against the linear slosh model.
    (a) The free surface at three times of the slide of $\movetime = \qty{1.05}{\s}$ at the fill depth \qty{8.5}{\cm}.
    (b) Height at the wall at \qty{8.5}{\cm} for $\movetime = \qty{0.83}{\s}$ (solid), which spills, and \qty{1.05}{\s} (dashed), simulated on \qty{2.5}{\mm} cells (black) and modeled (blue), with the rim (red) and the end of each slide (gray marks).
    (c) Peak height at the wall against $\movetime$, modeled (lines) and simulated (dots) at the fill depths $7.5$, $8$, and \qty{8.5}{\cm} (light to dark), with runs stopped near the top of the domain, whose peaks are lower bounds (triangles), runs on \qty{1.25}{\mm} cells at \qty{8}{\cm} (black squares), and the rim margin of each depth (dashed).}
    \label{fig: results slosh fluid}
\end{figure}

\subsubsection{Results}
The zonotent bound was at least the Monte Carlo estimate for all $282$ plans (\Cref{fig: results slosh}), by a factor of $1.6$ to $2.5$ where the estimate lay between $10^{-3}$ and $0.1$.
For the MJ plan with $\spillrisk = 0.01$, it was $0.0088$ against $0.0037$.
The union bound in place of \Cref{eq: first passage} would have been $2.4$ times larger.
On the grid of \qty{0.01}{\s} (\Cref{tab: results slosh}), the plans of the zonotent bound were one or two steps shorter than those of the Gaussian tail, one step longer than those of the Gaussian with the first-passage bound, and one or two steps longer than those of Monte Carlo.
Worst-case reachability accepted no plan, since its support reached at least $3.0$ times the threshold.
The plans of the nominal check, of $0.83$ and \qty{0.77}{\s}, spilled with probabilities of $0.9999$ and $1.0$ at \qty{8.5}{\cm}.

On cells of \qty{1.25}{\mm}, the fluid simulation and the model agreed on whether the water rose above the rim in $27$ of the $30$ runs (\Cref{fig: results slosh fluid}), with first crossings within \qty{0.016}{\s}.
In the other three, the model's peak lay $0.4$ to \qty{2.4}{\mm} below the rim margin while the simulated water crossed it late in the slide, so the longest slide that rose above the rim was at most \qty{0.04}{\s} longer in the simulation than in the model.
Where the model's peak was at most \qty{19}{\mm}, the simulated peak was $0.93$ to $1.04$ times the model's.
Above \qty{19}{\mm}, it was $1.06$ to $1.12$ times the model's.
On cells of \qty{2.5}{\mm}, the simulation and the model agreed in all $33$ runs, with simulated peaks $7$ to $11\%$ below the model's in the runs that stayed below the rim.

\subsubsection{Discussion}
Both hypotheses on the bound hold, with the gain over the Gaussian tail coming from the first-passage sets of \Cref{cor: first passage} alone.
The bound accepted plans where worst-case reachability accepted none and the plans of the nominal check almost surely spilled, at the cost of one grid step against the first-passage Gaussian, which is no bound.
The methods differ little because the probability of spilling falls tenfold per \qty{0.05}{\s} of $\movetime$.
The support enters this bound through the mismatch sets of \Cref{eq: slosh mismatch set}, which a Gaussian cannot give.

The hypothesis on the slosh model holds on the coarser grid, and on the finer one away from the spill boundary.
Near the boundary, the finer simulation spilled in slightly longer slides than the model, since the model's error in the peak grows with the amplitude, with a sign that depends on the grid.
The noise-free peaks of the zonotent bound's MJ plans are $74\%$ and $81\%$ of the smallest rim margin, where the model is accurate.
A smaller rim margin or a set of slosh models would extend the bound to the model's error near the rim.

\section{Conclusion}\label{sec: conclusion}
This paper introduces zonotents, which pair a zonotope with a density supported on it, so that one object bounds every reachable state and gives the probability of any subset.
Gaussian zonotents propagate exactly through affine maps and, by moment matching, approximately through Minkowski sums.
Our main contribution, probabilistic order reduction, reduces their supports in outer, inner, and free forms.
In the experiments, the free reduction met its budget in up to ten dimensions, and the Gaussian zonotent matched the accuracy of the untruncated Gaussian while, with uniform noise, coming much closer to simulation than the bounds of probabilistic zonotopes.
Inner reductions of a Gaussian overbound certified for the true system a gap that the exact reachable zonotope blocked, and the same overbound bounded the probability of spilling a cup of water in a linear slosh model.

\subsubsection*{Limitations}
We note four limitations.
First, our moment-matched Minkowski sum carries no guarantee and fits flat-topped convolutions, such as those of uniform noise, poorly (\Cref{subsec: results minkowski,subsec: results probzono}), whereas the guaranteed alternative, a Gaussian overbound (\Cref{prop: first exit}), loosens by a factor at every step that stays near one only for noise close to a truncated Gaussian.
Second, the bounds of the reductions (\Cref{prop: reduction guarantee,prop: free reduction,prop: free reduction share}) hold relative to the zonotent being reduced and only through exact operations, so only simulation checks the probabilities that use the moment-matched sum, while the narrow gap and the spill bounds hold for the true system.
Third, the Gaussian mass of \Cref{prop: gaussian mass} grows combinatorially with the dimension, so the inner reductions ran only in low dimensions, and they missed their budgets when the state was close to uniform on its support (\Cref{subsec: results collision,subsec: results reduction dimension}), where the free reduction, which needs no Gaussian mass of a zonotope, can replace them.
Fourth, our results for now only consider linear discrete-time dynamics (we anticipate extending to nonlinear systems with standard zonotope techniques).
Our slosh experiment therefore uses a linear slosh model, which underpredicts the simulated fluid peak near the rim (\Cref{subsec: results slosh}).

\section*{AI Usage Acknowledgment}
Anthropic's Claude Code, primarily Opus 5.5 High Effort was used extensively in preparing this paper.
The author drafted the text, conceived the core mathematical ideas, research questions, claims, experiments, and baselines.
Claude also drafted the text (especially proofs, which were checked by the author), implemented the code, suggested the moment matched Minkowski sum after the author experimented with iterative/nonlinear fit matching methods.
The author takes responsibility for all content.
In the spirit of openness, the author is making the Claude Skills used for this paper public:
\url{https://github.com/skousik/claude-skills}

\renewcommand{\bibfont}{\normalfont\footnotesize}
{\renewcommand{\markboth}[2]{}%
\printbibliography}

\end{document}